\documentclass{article}
 \pdfoutput=1
\usepackage{authblk} 
\usepackage{cite} 

\usepackage{graphicx}
\usepackage{amsmath,amsfonts}
\usepackage[margin=1in]{geometry}
\usepackage{array}
\usepackage{caption}
\usepackage{tikz}
\usepackage{listings}
\usetikzlibrary{snakes}
\usetikzlibrary{plotmarks}

\def\begeq{\begin{equation}}
\def\endeq{\end{equation}}

\def\begeqar{\begin{eqnarray}}
\def\endeqar{\end{eqnarray}}

\newcommand{\RI}{
\begin{tikzpicture}[scale=0.5]
\draw [black, line width=0.2] (0,-1) -- (1,0); 
\draw [black, line width=0.2] (0,1) -- (1,0); 
\draw [black, line width=0.2] (-1,0) -- (0,-1); 
\draw [black, line width=0.2] (-1,0) -- (0,1);  
\node (1) at (0,-1.5) {$\rho_1$};
\end{tikzpicture}}
\newcommand{\RII}{
\begin{tikzpicture}[scale=0.5]
\draw [black, line width=0.2] (0,-1) -- (1,0); 
\draw [black, line width=0.2] (0,1) -- (1,0); 
\draw [black, line width=0.2] (-1,0) -- (0,-1); 
\draw [black, line width=0.2] (-1,0) -- (0,1);  
\draw[black, line width=0.3mm, rounded corners=7pt] (-0.5,-0.5) -- (-0.1,0.0) -- (-0.5,0.5);
\node (1) at (0,-1.5) {$\rho_2$};
\end{tikzpicture}}
\newcommand{\RIII}{
\begin{tikzpicture}[scale=0.5]
\draw [black, line width=0.2] (0,-1) -- (1,0); 
\draw [black, line width=0.2] (0,1) -- (1,0); 
\draw [black, line width=0.2] (-1,0) -- (0,-1); 
\draw [black, line width=0.2] (-1,0) -- (0,1);  
\draw[black, line width=0.3mm, rounded corners=7pt] (0.5,-0.5) -- (0.1,0.0) -- (0.5,0.5);
\node (1) at (0,-1.5) {$\rho_3$};
\end{tikzpicture}}
\newcommand{\RIV}{
\begin{tikzpicture}[scale=0.5]
\draw [black, line width=0.2] (0,-1) -- (1,0); 
\draw [black, line width=0.2] (0,1) -- (1,0); 
\draw [black, line width=0.2] (-1,0) -- (0,-1); 
\draw [black, line width=0.2] (-1,0) -- (0,1);  
\draw[black, line width=0.3mm, rounded corners=7pt] (-0.5,-0.5) -- (0,-0.1) -- (0.5,-0.5);
\node (1) at (0,-1.5) {$\rho_4$};
\end{tikzpicture}}
\newcommand{\RV}{
\begin{tikzpicture}[scale=0.5]
\draw [black, line width=0.2] (0,-1) -- (1,0); 
\draw [black, line width=0.2] (0,1) -- (1,0); 
\draw [black, line width=0.2] (-1,0) -- (0,-1); 
\draw [black, line width=0.2] (-1,0) -- (0,1);  
\draw[black, line width=0.3mm, rounded corners=7pt] (-0.5,0.5) -- (0.,0.1) -- (0.5,0.5);
\node (1) at (0,-1.5) {$\rho_5$};
\end{tikzpicture}}
\newcommand{\RVI}{
\begin{tikzpicture}[scale=0.5]
\draw [black, line width=0.2] (0,-1) -- (1,0); 
\draw [black, line width=0.2] (0,1) -- (1,0); 
\draw [black, line width=0.2] (-1,0) -- (0,-1); 
\draw [black, line width=0.2] (-1,0) -- (0,1);  
\draw[black, line width=0.3mm, rounded corners=7pt] (-0.5,-0.5) -- (0.5,0.5);
\node (1) at (0,-1.5) {$\rho_6$};
\end{tikzpicture}}
\newcommand{\RVII}{
\begin{tikzpicture}[scale=0.5]
\draw [black, line width=0.2] (0,-1) -- (1,0); 
\draw [black, line width=0.2] (0,1) -- (1,0); 
\draw [black, line width=0.2] (-1,0) -- (0,-1); 
\draw [black, line width=0.2] (-1,0) -- (0,1);  
\draw[black, line width=0.3mm, rounded corners=7pt] (-0.5,0.5)  -- (0.5,-0.5);
\node (1) at (0,-1.5) {$\rho_7$};
\end{tikzpicture}}
\newcommand{\RVIII}{
\begin{tikzpicture}[scale=0.5]
\draw [black, line width=0.2] (0,-1) -- (1,0); 
\draw [black, line width=0.2] (0,1) -- (1,0); 
\draw [black, line width=0.2] (-1,0) -- (0,-1); 
\draw [black, line width=0.2] (-1,0) -- (0,1);  
\draw[black, line width=0.3mm, rounded corners=7pt] (-0.5,-0.5) -- (-0.1,0.0) -- (-0.5,0.5);
\draw[black, line width=0.3mm, rounded corners=7pt] (0.5,-0.5) -- (0.1,0.0) -- (0.5,0.5);
\node (1) at (0,-1.5) {$\rho_8$};
\end{tikzpicture}}
\newcommand{\RIX}{
\begin{tikzpicture}[scale=0.5]
\draw [black, line width=0.2] (0,-1) -- (1,0); 
\draw [black, line width=0.2] (0,1) -- (1,0); 
\draw [black, line width=0.2] (-1,0) -- (0,-1); 
\draw [black, line width=0.2] (-1,0) -- (0,1);  
\draw[black, line width=0.3mm, rounded corners=7pt] (-0.5,-0.5) -- (0,-0.1) -- (0.5,-0.5);
\draw[black, line width=0.3mm, rounded corners=7pt] (-0.5,0.5) -- (0.,0.1) -- (0.5,0.5);
\node (1) at (0,-1.5) {$\rho_9$};
\end{tikzpicture}}
\newcommand{\arrowaup}{
\begin{tikzpicture}[scale=0.5]
\draw [black, line width=0.2] (0,-1) -- (1,0); 
\draw [black, line width=0.2] (0,1) -- (1,0); 
\draw [black, line width=0.2] (-1,0) -- (0,-1); 
\draw [black, line width=0.2] (-1,0) -- (0,1);  
\draw[->,>=latex, black, line width=0.3mm] (-0.5,-0.5) -- (0,0);
\draw[black, line width=0.2mm, dashed, dash pattern=on 1pt off 1pt] (0,0) -- (0.5,0.5);
\draw[black, line width=0.2mm, dashed, dash pattern=on 1pt off 1pt] (0.5,-0.5) -- (-0.5,0.5);
\node (1) at (0,-1.5) {$\alpha^{-1}$};
\end{tikzpicture}}
\newcommand{\arrowadown}{
\begin{tikzpicture}[scale=0.5]
\draw [black, line width=0.2] (0,-1) -- (1,0); 
\draw [black, line width=0.2] (0,1) -- (1,0); 
\draw [black, line width=0.2] (-1,0) -- (0,-1); 
\draw [black, line width=0.2] (-1,0) -- (0,1);  
\draw[<-,>=latex, black, line width=0.3mm] (-0.5,-0.5) -- (0,0);
\draw[black, line width=0.2mm, dashed, dash pattern=on 1pt off 1pt] (0,0) -- (0.5,0.5);
\draw[black, line width=0.2mm, dashed, dash pattern=on 1pt off 1pt] (0.5,-0.5) -- (-0.5,0.5);
\node (1) at (0,-1.5) {$\alpha$};
\end{tikzpicture}}
\newcommand{\arrowdup}{
\begin{tikzpicture}[scale=0.5]
\draw [black, line width=0.2] (0,-1) -- (1,0); 
\draw [black, line width=0.2] (0,1) -- (1,0); 
\draw [black, line width=0.2] (-1,0) -- (0,-1); 
\draw [black, line width=0.2] (-1,0) -- (0,1);  
\draw[->,>=latex, black, line width=0.3mm] (0,0) -- (0.5,0.5);
\draw[black, line width=0.2mm, dashed, dash pattern=on 1pt off 1pt] (-0.5,-0.5) -- (0,0);
\draw[black, line width=0.2mm, dashed, dash pattern=on 1pt off 1pt] (0.5,-0.5) -- (-0.5,0.5);
\node (1) at (0,-1.5) {$\alpha$};
\end{tikzpicture}}
\newcommand{\arrowddown}{
\begin{tikzpicture}[scale=0.5]
\draw [black, line width=0.2] (0,-1) -- (1,0); 
\draw [black, line width=0.2] (0,1) -- (1,0); 
\draw [black, line width=0.2] (-1,0) -- (0,-1); 
\draw [black, line width=0.2] (-1,0) -- (0,1);  
\draw[<-,>=latex, black, line width=0.3mm] (0,0) -- (0.5,0.5);
\draw[black, line width=0.2mm, dashed, dash pattern=on 1pt off 1pt] (-0.5,-0.5) -- (0,0);
\draw[black, line width=0.2mm, dashed, dash pattern=on 1pt off 1pt] (0.5,-0.5) -- (-0.5,0.5);
\node (1) at (0,-1.5) {$\alpha^{-1}$};
\end{tikzpicture}}
\newcommand{\turnacup}{
\begin{tikzpicture}[scale=0.5]
\draw [black, line width=0.2] (0,-1) -- (1,0); 
\draw [black, line width=0.2] (0,1) -- (1,0); 
\draw [black, line width=0.2] (-1,0) -- (0,-1); 
\draw [black, line width=0.2] (-1,0) -- (0,1);  
\draw[->,>=latex, black, line width=0.3mm, rounded corners=7pt] (-0.5,-0.5) -- (0,0) -- (-0.5,0.5);
\draw[black, line width=0.2mm, dashed, dash pattern=on 1pt off 1pt] (0.5,-0.5) -- (0,0) -- (0.5,0.5);
\node (1) at (0,-1.5) {$\beta$};
\end{tikzpicture}}
\newcommand{\turnacdown}{
\begin{tikzpicture}[scale=0.5]
\draw [black, line width=0.2] (0,-1) -- (1,0); 
\draw [black, line width=0.2] (0,1) -- (1,0); 
\draw [black, line width=0.2] (-1,0) -- (0,-1); 
\draw [black, line width=0.2] (-1,0) -- (0,1);  
\draw[<-,>=latex, black, line width=0.3mm, rounded corners=7pt] (-0.5,-0.5) -- (0,0) -- (-0.5,0.5);
\draw[black, line width=0.2mm, dashed, dash pattern=on 1pt off 1pt] (0.5,-0.5) -- (0,0) -- (0.5,0.5);
\node (1) at (0,-1.5) {$\beta^{-1}$};
\end{tikzpicture}}
\newcommand{\turnbdup}{
\begin{tikzpicture}[scale=0.5]
\draw [black, line width=0.2] (0,-1) -- (1,0); 
\draw [black, line width=0.2] (0,1) -- (1,0); 
\draw [black, line width=0.2] (-1,0) -- (0,-1); 
\draw [black, line width=0.2] (-1,0) -- (0,1);  
\draw[->,>=latex, black, line width=0.3mm, rounded corners=7pt] (0.5,-0.5) -- (0,0) -- (0.5,0.5);
\draw[black, line width=0.2mm, dashed, dash pattern=on 1pt off 1pt] (-0.5,-0.5) -- (0,0) -- (-0.5,0.5);
\node (1) at (0,-1.5) {$\beta^{-1}$};
\end{tikzpicture}}
\newcommand{\turnbddown}{
\begin{tikzpicture}[scale=0.5]
\draw [black, line width=0.2] (0,-1) -- (1,0); 
\draw [black, line width=0.2] (0,1) -- (1,0); 
\draw [black, line width=0.2] (-1,0) -- (0,-1); 
\draw [black, line width=0.2] (-1,0) -- (0,1);  
\draw[<-,>=latex, black, line width=0.3mm, rounded corners=7pt] (0.5,-0.5) -- (0,0) -- (0.5,0.5);
\draw[black, line width=0.2mm, dashed, dash pattern=on 1pt off 1pt] (-0.5,-0.5) -- (0,0) -- (-0.5,0.5);
\node (1) at (0,-1.5) {$\beta^{-1}$};
\end{tikzpicture}}

\title{A new look at the collapse of two-dimensional polymers}
\author[1,2]{\'Eric Vernier}
\author[1,3]{Jesper Lykke Jacobsen}
\author[2,4]{Hubert Saleur}
\affil[1]{LPTENS, \'Ecole Normale Sup\'erieure, 24 rue Lhomond, 75231 Paris, France}
\affil[2]{IPhT, CEA Saclay, 91191 Gif-sur-Yvette, France}
\affil[3]{Universit\'e Pierre et Marie Curie, 4 place Jussieu, 75252 Paris, France}
\affil[4]{USC Physics Department, Los Angeles CA 90089, USA}
\date{March 30, 2015}

\begin{document}
\maketitle

\begin{abstract}

We study the collapse of two-dimensional polymers, via an O($n$) model on the square lattice
that allows for dilution, bending rigidity and short-range monomer attractions. This model contains
two candidates for the theta point, $\Theta_{\rm BN}$ and $\Theta_{\rm DS}$, both exactly solvable.
The relative stability of these points, and the question of which one describes the `generic' theta point,
have been the source of a long-standing debate. Moreover, the analytically predicted exponents of
$\Theta_{\rm BN}$ have never been convincingly observed in numerical simulations.

In the present  paper, we shed a new light on this confusing situation.
We show in particular that   the continuum limit of $\Theta_{\rm BN}$ is an unusual conformal field theory, made in fact of a simple
dense polymer decorated with {\sl non-compact degrees of freedom}. This implies in particular
that the critical exponents take continuous rather than discrete values, and that corrections to scaling
lead to an unusual integral form. Furthermore, discrete states may emerge from the continuum, but
the latter are only normalizable---and hence observable---for appropriate values of the model's parameters.
We check these findings numerically. We also probe the non-compact degrees
of freedom in various ways, and establish that they are related to fluctuations of the density of monomers.
Finally, we construct a field theoretic model of the vicinity of $\Theta_{\rm BN}$ and examine the flow along
the multicritical line between $\Theta_{\rm BN}$ and $\Theta_{\rm DS}$.

\end{abstract}

\section{Introduction}

Two-dimensional self-avoiding walks with additional generic short-range attraction are believed to experience a collapse transition
as the temperature is lowered \cite{DJ76,deGennes}. The high-temperature phase is in the universality class of ordinary self-avoiding walks (SAWs), also known
as {\em dilute polymers}, with well-known exponents. The low-temperature phase is in the universality class of so-called
{\em dense polymers}, also with well-known exponents \cite{DS-NPB-87}.%
\footnote{The zero-temperature case, where the walks are fully packed, is known as Hamiltonian walks. It presents more subtle
features and strong non-universality, with critical exponents that are lattice-dependent \cite{BN-FPLhex-94,KGN-FPLhex-96,BBN-FPLsq-96,JK-FPLsq-98}
and depend continuously on the stiffness for models of semi-flexible walks \cite{JK-semiflex-04,JA-semiflex-09}. We will not discuss this further here.}
In between sits the so-called {\em theta point}, where the polymers are somewhat more compact than in the dilute phase,
but still have a fractal dimension strictly smaller than two.

The theta point is well identified in the language of the O($n$) field theory (in the polymer limit $n\to 0$) as a tricritical point \cite{DJ76,deGennes},
while the ordinary SAWs correspond to the usual critical point. Non-generic attraction between monomers would lead to higher criticality
for the transition point, leading to what is sometimes called a {\em theta prime} point \cite{Vander,PCJS}. 

It is unfortunately  difficult to  identify  `the'  theta point based on its expected tricritical nature. When dealing with geometrical problems, the counting of physical observables---many of which are, in a sense, non-local--- is ambiguous. Moreover, in two dimensions, the Landau-Ginzburg picture is hardly able to organise the zoo of known universality classes, which is rendered even more complicated by the lack of unitarity inherent to geometrical problems. Finally, producing reliable numerical results turns out unexpectedly hard.  This means the identification of the 
theta point  critical exponents has led to a long-standing controversy.

In 1987, Duplantier and Saleur \cite{DS87} were able to solve exactly a model (first proposed by Coniglio et al.\ \cite{Coniglio}) of SAWs on the honeycomb
lattice with a particular type of attractive interaction. They conjectured that this interaction was generic enough to put the model in the theta universality class,
and obtained the exponents
\begin{eqnarray}
 \nu_\Theta &=& {4\over 7} \approx 0.571 \,, \nonumber\\
 \gamma_\Theta &=& {8\over 7} \approx 1.143 \,, \nonumber\\
 \phi_\Theta &=& {3\over 7} \approx 0.429 \,.\label{expoDS}
\end{eqnarray}
Later, the same authors \cite{DS89} carried out extensive numerical simulations to explore the stability of their conjectured theta point against additional
attractions, and concluded that indeed, the exponents they had obtained described the generic (tricritical) theta point. These exponents were also found
in agreement with numerical simulations of interacting self-avoiding walks on other lattices \cite{DerrSal},
but it is fair to say that the proximity of the dense phase with
very strong corrections to scaling rendered these results a little less definite than one would have liked. For the square lattice and presumably generic
nearest-neighbour attractions, the best recent numerical results are \cite{Cara}
\begin{eqnarray}
 \nu_{\Theta}^{\rm num} &=& 0.570(2) \,, \nonumber\\
 \phi_{\Theta}^{\rm num} &=& 0.46(3) \,.
\end{eqnarray}

Meanwhile, in a beautiful series of papers \cite{Nien1,Nien2,Nien3,Nien4}, Nienhuis and collaborators managed to find an exactly solvable O($n$) model
on the square lattice with a rich set of phase transitions. One of these has, for $n \to 0$, all the required characteristics of the theta point, but leads, disturbingly,
to critical exponents%
\footnote{The subscript BN refers to Bl\"ote and Nienhuis \cite{Nien1} and will be used extensively throughout this work.}
which are different from those of \cite{DS87}:
\begin{eqnarray}
\nu_{\rm BN} &=& {12\over 23}\approx 0.522 \,, \nonumber\\
\gamma_{\rm BN} &=& {53\over 46}\approx 1.152 \,. 
\label{exponien}
\end{eqnarray}
It was suggested at the time in \cite{Nien4} that this solvable point might be in fact  `the' generic theta point, implying  that the point obtained by Duplantier and Saleur \cite{DS87}
was, in fact, a higher multicritical point - although  it is hard to see how  the exponent $\nu$ in (\ref{expoDS}) for a higher multicritical point could be larger than the one (\ref{exponien}) for the generic theta point.

To complicate the matters further, the exponents obtained in \cite{Nien4}, despite the impeccable derivation obtained via the Bethe Ansatz, have never
quite been observed in numerical simulations. Frustratingly, for the very model discussed in \cite{Nien4}, and using the most state-of-the-art numerical
techniques available today, the exponents obtained in direct simulations \cite{Bedini} read
\begin{eqnarray}
 \nu^{\rm num}_{\rm BN} &=& 0.576(6) \,, \nonumber\\
 \gamma^{\rm num}_{\rm BN} &=&1.045(5) \,.
\end{eqnarray}
and are, bizarrely, close to the ones (\ref{expoDS}) obtained by Duplantier Saleur! Similar conclusions were drawn in \cite{FosterDep12} in the context of the measure of surface critical exponents, both from transfer matrix and DMRG calculations.  

The discrepancy with respect to the analytical results (\ref{exponien}) is essentially unheard of in the field of exactly solvable models and conformal field
theory in two dimensions, where, usually, theory and numerical experiments match almost perfectly.%
\footnote{To give but one example of an (at least seemingly) very comparable situation, the exponent $\gamma$ for Hamiltonian walks on the square
lattice was found numerically to be \cite{BBN-FPLsq-96} $\gamma = 1.0444(1)$, in perfect agreement with the subsequent analytical solution
\cite{JK-FPLsq-98} $\gamma = \frac{117}{112} \approx 1.0446$.}
Of course, one important point is that in  \cite{Nien4}, the quantities which are studied involve a grand canonical ensemble of walks where the length
fluctuates, while in the simulations of \cite{Bedini} polymers have a fixed length, and the ensemble is canonical. However, this point is usually easily
taken care of by considerations of Legendre transform, and has never, in all the other cases studied so far, led to any particular difficulty. 

While evidence gathered over the last many years (see, e.g., \cite{Lee,Gherardi} for recent contributions) has confirmed that the exponents
obtained in \cite{DS87} indeed are those of the generic theta point, the meaning of the exactly solved model in \cite{Nien4} has remained totally
unclear: what is its universality class in the O($n$) field theory language? Why are the exponents so hard to observe numerically?

The detailed answer to this question is the object of this paper and turns out to be a rather long story. It can however be summarised rather succinctly:
The universality class of the model in \cite{Nien4} is profoundly different from the one of ordinary SAWs. While for the latter, the spectrum of critical
exponents is {\em discrete}---as is the case for most familiar models%
\footnote{A few noticeable exceptions have nevertheless appeared in the recent literature \cite{IkhlefAFPotts,IkhlefBH,VJS1,VernierPotts}.}
of statistical mechanics---for the former, the critical exponents form in fact a {\em continuum}, of which the values (\ref{exponien}) obtained in \cite{Nien4}
are simply the lower bounds. While, as we shall see below, this explains the numerical difficulties encountered so far, it also shows that the continuum
field theoretic description is profoundly different from the O($n \to 0$) model $\Phi^6$ theory expected for the theta point. The field theoretic description
instead involves a non-compact target, and is related with the {\em Black Hole sigma model} conformal field theory that has been discussed intensively
in the string theory literature \cite{BlackHoleCFT,Troost,RibSch}.

\section{The polymer model and its phase diagram}
\label{sec:2}

The model considered in \cite{Nien4}  is one of a single polymer on the square lattice, where edges can be visited at most once. The polymer is not allowed to cross itself, but
collisions where two pieces of the polymer barely avoid each other are allowed at all sites: an energy $-\epsilon_t$ is associated to this in order to
take into account the presence of medium-range attractions in the physics of the theta point (see figure \ref{fig:VISAWconfig}).

Additionally, it is convenient to also allow for some stiffness, and thus associate with two parallel consecutive monomers an energy $-\epsilon_s$.
The resulting object is called a semi-flexible VISAW (vertex-interacting self-avoiding walk), in the notations of \cite{Bedini} that we follow here.
The partition function of a polymer made of $N$ monomers is thus given by
\begin{equation}
Z_N(\tau,p)=\sum_{\rm VISAW} \tau^{\scriptsize{\hbox{number of doubly visited sites}}}~~ p^{\scriptsize{\hbox{number of straight segments}}} \,,
\end{equation}
with $\tau\equiv e^{\beta \epsilon_t}$ and $p\equiv e^{\beta\epsilon_s}$.
While Monte Carlo simulations focus on polymers of fixed large length $N$, the integrable model deals instead with a grand canonical ensemble
where the monomers have a fugacity  $K$, so the partition function is 
\begin{equation}
 G=\sum_{N=0}^\infty K^N Z_N(\tau,p) \,.
\end{equation}
\begin{figure}
\begin{center}
\begin{tikzpicture}[scale=0.7]
\foreach \x in {0,1,2,3,4,5,5}
{ 
\draw[black, line width=0.2mm] (\x,0) -- (\x,5);
}
 \foreach \y in {0,1,2,3,4,5,5}
{ 
\draw[black, line width=0.2mm] (0,\y) -- (5,\y);
}
\draw[black, line width=1.0mm, rounded corners=7pt] (1,0) -- (1,1) -- (1,2) -- (2,2) -- (2,3) -- (3,3) -- (3,2) -- (4,2) -- (4,3) -- (3,3) -- (3,4) -- (4,4) -- (4,3) -- (5,3);
\draw[red] (1,1) ellipse (0.3 and 0.5);
\draw[blue] (3,3) ellipse (0.5 and 0.5);
 \node[red] at (1.4,1.3) {$p$};
  \node[black] at (1.5,2.3) {$K$};
   \node[blue] at (2.5,3.5) {$\tau$};
\end{tikzpicture}
\end{center}
\caption{Configuration of the semi-flexible vertex-interacting self-avoiding walk (VISAW).}
\label{fig:VISAWconfig}
\end{figure}
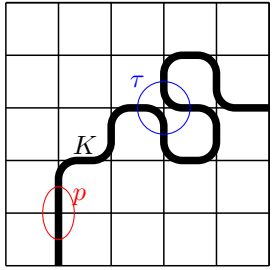

\begin{figure}
  \begin{center}
  \begin{tikzpicture}[scale=4.5]    
     \draw (1,0,0)--(1,1,0)--(0,1,0); 
   \draw[dashed] (0,1,0)--(0,0,0)--(1,0,0); 
\draw (0,0,1)--  node[below] {\Large $p$} (1,0,1);
\draw[->,>=latex,very thick]  (0.49,0,1) -- (0.5,0,1);
\draw[->,>=latex,very thick]  (0.,0.49,1) -- (0.,0.5,1);
\draw[->,>=latex,very thick]  (0.,0,0.51) -- (0.,0,0.5);
\draw(1,0,1)--(1,1,1)--(0,1,1)-- node[left] {\Large $\tau$} (0,0,1);
\draw[dashed] (0,0,0) --  node[above] {\Large $K$} (0,0,1); 
\draw (1,0,0) -- (1,0,1); 
\draw (1,1,0) -- (1,1,1); 
\draw (0,1,0) -- (0,1,1); 

\foreach[count=\xi] \v in {0,0.02,...,1.02} {
\draw[red, rounded corners=20pt] (\v,0.1-0.1*\v,0+0.05*\v) -- (\v,0.3-0.1*\v,0.5+0.05*\v) -- (\v,0.5-0.1*\v,0.7+0.05*\v);
\draw[blue!\xi]  (\v,0.5-0.1*\v,0.7+0.05*\v) --  (\v,0.5,0.);
\draw[black, dashed] (\v,0.5-0.1*\v,0.7+0.05*\v) -- (\v,1,0.9);
}
\draw[violet, line width=3pt] (0,0.5,0.7) -- (1,0.4,0.75);
 \fill[orange] (0,0.5,0.7) circle [radius=0.8pt];
 \draw (0,0.5,0.7+0.025) node[below] {$\Theta_{\rm DS}$}; 
 \fill[green] (0.5,0.5-0.05,0.7+0.025) circle [radius=0.8pt];   
 \draw (0.5,0.5-0.05,0.7+0.055) node[below] {$\Theta_{\rm BN}$}; 
 
 \draw[violet, very thick, ->,>=latex]  (-0.2,0.2,1) -- (0.3,0.5-0.03,0.7+0.06); 
     \node[left] at  (-0.2,0.2,1) {multicritical line};
       \draw[red, very thick, ->,>=latex]  (1.1,-0.1,0) -- (0.55,0.25,0.52); 
       \node[right] at  (1.1,-0.1,0) {dilute};
              \draw[very thick, ->,>=latex]  (1.1,-0.3,0) -- (0.55,0.1,0.52);  
       \node[right] at  (1.1,-0.3,0) {massive};
       \draw[blue, very thick, ->,>=latex]  (1.1,0.5,0) -- (0.95,0.4,0.52);    
       \node[right] at  (1.1,0.5,0) {Ising ordering};
  \draw[purple, very thick, ->,>=latex]  (1.1,0.2,0) -- (0.98,0.27,0.35);  
      \node[right] at  (1.1,0.2,0) {dense (Ising disordered)};
  \draw[purple, very thick, ->,>=latex]  (1.1,0.8,0) -- (0.98,0.7,0.35);  
      \node[right] at  (1.1,0.8,0) {dense (Ising ordered)};
  \end{tikzpicture}
  \caption{Schematic phase diagram of the two-dimensional VISAW. The nature of the different phases and transitions between them is explained in the main text.
   The two candidates, $\Theta_{\rm DS}$ and $\Theta_{\rm BN}$, for the theta point universality class are represented, respectively, by an orange (DS) and a
   green (BN) dot.}
  \label{fig:phasediagram}
\end{center}
\end{figure}
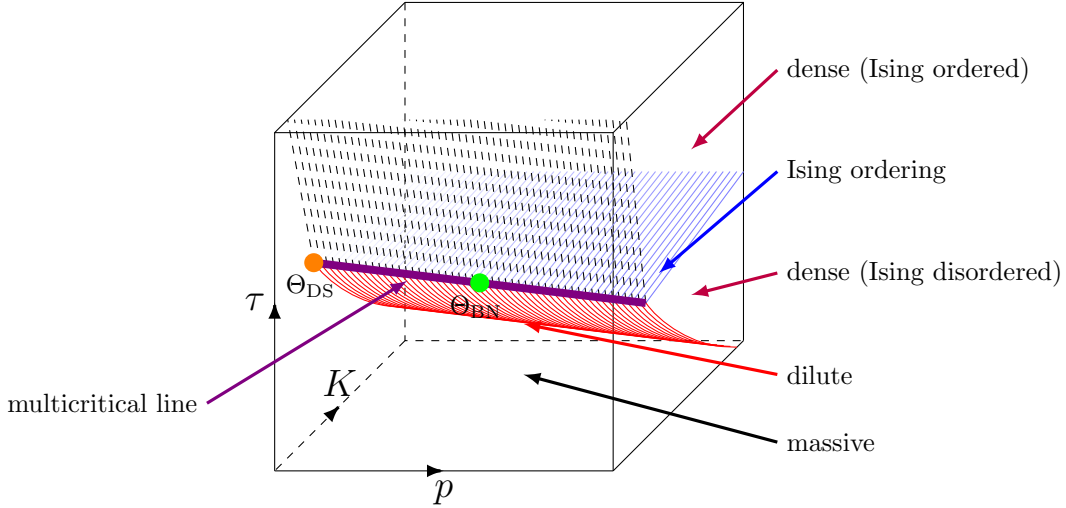 

The phase diagram in the $(p,K,\tau)$ parameter space, which has been the object of several numerical studies \cite{Blotenum1,Blotenum2,Blotenum3},
is depicted schematically in figure \ref{fig:phasediagram}.
For small values of the interaction parameter $\tau$, the physics is essentially the same as for usual (non-interacting) SAWs: for small values of the fugacity $K$
the polymer is in a massive phase and remains of finite total length as the size of the system is sent to infinity, whereas for high values of $K$ the polymer is in a
critical, {\em dense} phase where it covers a finite fraction of the lattice edges in the thermodynamic limit; the two phases are separated at intermediate $K$ by
a different, {\em dilute} critical phase (represented by a red surface in figure \ref{fig:phasediagram}).

In order to investigate the effect of larger values of $\tau$, it is instructive to consider Ising degrees of freedom living on the faces of the square lattice, such that
two Ising spins on adjacent faces share the same orientation iff they are separated by a monomer. While this Ising degree of freedom is disordered in the usual
SAW dense phase, it is clearly ordered in the 
so-called {\em completely packed} (Eulerian walk) limit of large $K$ and $\tau$, and therefore some critical Ising surface should be expected in the phase
diagram (shown in blue in figure \ref{fig:phasediagram}). Above this critical surface, the transition between the massive and dense (ordered) phases is now of first
order, and is represented by a dashed surface in the figure.

Whereas all the phases described so far are by now well understood,%
\footnote{Note however that the $p=0$ plane reveals some surprises for generic $n$ \cite{Blotenum3} that we plan to unravel in a future piece of work \cite{p0unpub}.}
this is not the case of the multicritical line joining the dilute, Ising critical, and first-order
surfaces, and which contains both candidates under consideration for the theta point physics:
\begin{itemize}
 \item The integrable point $\Theta_{\rm BN}$ \cite{Nien4}, depicted as a green dot in figure \ref{fig:phasediagram}, which will be our main focus throughout this paper.
 \item The universality class  of \cite{DS87} is known \cite{Nien4} to describe the 
 particular integrable point $\Theta_{\rm DS}$ situated at $p=0$, that we depict as an orange dot in figure \ref{fig:phasediagram}.
\end{itemize}

The precise location in the $(p,K,\tau)$ phase diagram of the points $\Theta_{\rm BN}$ and $\Theta_{\rm DS}$, as well as that of three further integrable points,
follows from the integrable solution \cite{Nien1} and various transformations, of which we shall review the necessary details in section~\ref{section:correspondenceloopvertex}.
Suffice it here to say that the integrable model can be formulated in terms of a loop model with the following nine different local configurations around each lattice site:
\begin{equation}
\RI \quad \RII \quad \RIII \quad \RIV \quad \RV \quad \RVI \quad \RVII \quad \RVIII \quad \RIX \label{vertices}
\end{equation}
The corresponding Boltzmann weights are $\rho_1,\ldots,\rho_9$ as indicated, and there is an additional non-local weight $n$ per loop. We set the loop weight $n=0$
to obtain polymers. The integrable model gives the same weights to local configurations that are related by a horizontal and/or vertical reflection, so that
$\rho_2 = \rho_3$ and $\rho_4 = \rho_5$. It contains a further (so-called spectral) parameter and various gauge factors $\pm 1$ that can be adjusted to obtain
isotropic solutions for which we have also $\rho_2 = \rho_4$ and $\rho_8 = \rho_9$. The correspondence between $(\rho_1,\rho_2,\rho_6,\rho_8)$ 
with the parameters used in figure~\ref{fig:phasediagram} is then easily seen to be
\begin{equation}
 p = \frac{\rho_6}{\rho_2} \,, \qquad
 K = \frac{\rho_2}{\rho_1} \,, \qquad
 \tau = \frac{\rho_1 \rho_8}{(\rho_2)^2} \,.
\end{equation}
We also note that \cite{Bedini} prefers to trade $\tau$ for another parameter defined by $w = K^2 \tau = \rho_8 / \rho_1$.

The integrable model studied in \cite{Nien1,Nien2,Nien3,Nien4} is defined in terms of trigonometric weights $(\rho_1,\ldots,\rho_9;n)$ whose periodicities are such that,
after constraining to isotropic $n=0$ cases, there are in fact four distinct solutions for the parameters $(p,K,\tau)$. These are referred to as branches 1, 2, 3, 4 in \cite{Nien1},
and as regimes I, II, III in \cite{Nien4} (regime I corresponds to two branches). The point $\Theta_{\rm BN}$ corresponds to regime III (or branch 3) and the weights are
\begin{eqnarray}
 p_{\rm BN} &=& \sqrt{2} \sin \left( \frac{\pi}{16} \right) \approx 0.275899 \,, \nonumber \\
 K_{\rm BN} &=& \left[ 2 \cos \left( \frac{\pi}{16} \right) \left( 1 + \frac{1}{\sqrt{2}} \tan \left( \frac{\pi}{16} \right) \right) \right]^{-1} \approx 0.446933 \,, \nonumber \\
 \tau_{\rm BN} &=& \frac12 \left( 2 + \sqrt{2} + \sqrt{2 + \sqrt{2}} \right) \approx 2.630986 \,. \label{polymerweights}
\end{eqnarray}
The other three solutions can be shown to correspond to \cite{Nien1} a point in the dense phase (above the blue surface in figure~\ref{fig:phasediagram});
a point at the Ising ordering transition (on the blue surface); and a point in the dilute phase (on the red surface in figure~\ref{fig:phasediagram}).%
\footnote{For completeness, we notice that there is another point in the dilute phase which is extremely well studied and whose position is conjecturally
exactly known. It is the critical point $K_{\rm c}$ of the combinatorial SAW model, with $p=1$ and $\tau = 0$, in which vertex-avoiding walks are simply
weighted by $K$ per monomer. This is well suited for exact enumeration studies, of which the most recent \cite{Clisby12} confirms to almost 12 decimal places
the validity of a twenty year old conjecture \cite{Conway93} that $K_{\rm c}$ is the positive real root of the polynomial $581 K^4 + 7 K^2 - 13$.}

The point $\Theta_{\rm DS}$ is derived from another---simpler---integrable model in which only $\rho_8$ and $\rho_9$ are non-zero.
We denote the loop weight in this completely packed model by $\tilde{n}$.
It is equivalent to the $Q=\tilde{n}^2$ state critical Potts model \cite{Baxter73}, and is referred to as branch 0 in \cite{Nien1}.
One can shown that it is equivalent to a model of the type (\ref{vertices}) with non-zero $\rho_1$ via a simple trick \cite{Nien1} that we now recall.
Indeed, consider a loop model with only vertices $\rho_8$ and $\rho_9$ and loop weight $\tilde{n} = n+1$.
Draw each loop independently in black and white colour, giving weight $n$ to black loops and $1$ to white loops. The trivial weight given 
to white loops means that we do not need to keep track of their number, so the colour white can be considered invisible. This produces 
a model of the type (\ref{vertices}) in which $\rho_6 = \rho_7 = 0$. Starting from the completely packed model with normalised, isotropic weights,
$\rho_8 = \rho_9 = 1$, we obtain a dilute model with $\rho_2=\rho_3=\rho_4=\rho_5 = 1$ and $\rho_1 = 2$. 
Setting in particular $n=0$ we finally obtain 
\begin{equation}
 p_{\rm DS} = 0 \,, \qquad
 K_{\rm DS} = \frac12 \,, \qquad
 \tau_{\rm DS} = 2 \,.
\end{equation}
By means of the $\tilde{n} = n+1$ mapping, this model is equivalent to percolation (the $Q=1$ state Potts model), just as the theta point model originally
exhibited in \cite{DS87}. In particular, the critical exponents at $\Theta_{\rm DS}$ coincide with those of (\ref{expoDS}).

\section{Relation with the O($n$) loop and Izergin-Korepin vertex models}
\label{section:correspondenceloopvertex}

An important fact---which is at the root of the exact solution in \cite{Nien4}---is that the point $\Theta_{\rm BN}$ is just the particular $n=0$ case
of a more general solvable loop model related to the Izergin-Korepin $\check{R}$-matrix. We now review this in some detail. 

The  Izergin-Korepin $\check{R}$-matrix (also known as the $a_2^{(2)}$ $\check{R}$-matrix) is a solution of the Yang-Baxter equations acting on
two copies of the spin-1 representation of $U_{\mathfrak{q}}(sl_2)$. It is most conveniently expressed in terms of the quantum group projectors $P_s$
on spins $s=0,1,2$:
\begin{equation}
 \check{R}^{(2)} \propto P_2  + \frac{\mathfrak{q}^4 x -1}{\mathfrak{q}^4 -x} P_1 +  \frac{\mathfrak{q}^6 x +1}{\mathfrak{q}^6 +x}P_0 \,.
 \label{R2}
\end{equation}
We will set in the following  $\mathfrak{q} = \mathrm{e}^{\mathrm{i} {\gamma \over 2}}$ and $x = \mathrm{e}^{2\lambda}$, in order to match the notations
of Ref.~\cite{GalleasMartins04}.

While the most natural context to interpret such an $\check{R}$-matrix is in terms of a 19-vertex model, it turns out that a few simple transformations allow
a reinterpretation in terms of a dilute loop model \cite{ZhouBatchelor}.%
\footnote{This is somewhat similar to what happens for the $a_1^{(1)}$ model $\check{R}$-matrix, which is both related with the 6-vertex model and with
a completely packed $O(n)$ loop model \cite{BKW76}.}
 
In order to discuss this---and in view of the large literature on the subject---it is useful to review normalisations and conventions in detail. First, let us fix
the normalisation of the $\check{R}$-matrix. We start from the matrix $\check{R}_{\rm GM}$ given in \cite{GalleasMartins04}, which is related to (\ref{R2}) by
\begin{equation}
 \check{R}_{\rm GM} = \mathrm{i} \frac{\left(q^2-x\right)\left(q^3+x\right)}{x q^{5 \over 2}} \check{R}^{(2)} \,,
\end{equation}
where $q=\mathfrak{q}^2$. We then define our model by
\begin{equation}
 \check{R} =  \frac{1}{2 \left(\sin \left(\frac{\gamma }{2}\right)-\sin \left(\frac{5 \gamma }{2}\right)\right)}  \check{R}_{\rm GM} \,.
\end{equation}

Next, we graphically represent the edges with $S_z=0$ by empty edges, and those with $S_z = \pm 1$ by oriented lines. 
We then apply the following gauge transformations on the matrix elements of $\check{R}$:
\begin{center}
 \arrowaup \quad \arrowadown \quad \arrowdup \quad \arrowddown
 \label{gaugetransf}
\end{center}
with $\alpha = \mathrm{i} \mathrm{e}^{\mathrm{i}\gamma- \lambda}$. Then perform the following transformation
\begin{center}
 \turnacup \quad \turnacdown \quad \turnbdup \quad \turnbddown
\end{center}
with $\beta = -\mathrm{i} \mathrm{e}^{-\mathrm{i}\gamma}$. 
Considering now the model as a gas of dilute loops, each loop carrying the sum over its two possible orientations, the local
configurations around a vertex are precisely those shown in (\ref{vertices}). The latter gauge transformation
amounts to assigning to every closed loop a weight
\begin{equation}
 n = \beta^2 + \beta^{-2} = -2 \cos(2\gamma) \,, \label{loopweight}
\end{equation}
and the weights $\rho_1,\ldots,\rho_9$ assigned to each vertex configuration are the following
\begin{eqnarray}
\rho_1&=&1+{\sin u_{\rm ZB} \sin(3\lambda_{\rm ZB}-u_{\rm ZB})
\over \sin2\lambda_{\rm ZB}\sin3\lambda_{\rm ZB}} \,, \nonumber\\
\rho_2&=&\rho_3={\sin(3\lambda_{\rm ZB}-u_{\rm ZB})\over\sin 3\lambda_{\rm ZB}} \,, \nonumber\\
\rho_4&=&\rho_5={\sin u_{\rm ZB} \over\sin3\lambda_{\rm ZB}} \,, \nonumber\\
\rho_6&=&\rho_7={\sin u_{\rm ZB}\sin(3\lambda_{\rm ZB}-u_{\rm ZB})\over \sin 2\lambda_{\rm ZB}\sin3\lambda_{\rm ZB}} \,, \nonumber\\
\rho_8&=&{\sin(2\lambda_{\rm ZB}-u_{\rm ZB})\sin(3\lambda_{\rm ZB}-u_{\rm ZB})\over \sin 2\lambda_{\rm ZB}\sin3\lambda_{\rm ZB}} \,, \nonumber\\
\rho_9&=&-{\sin u_{\rm ZB}\sin(\lambda_{\rm ZB}-u_{\rm ZB})\over \sin 2\lambda_{\rm ZB}\sin3\lambda_{\rm ZB}} \,, \nonumber \\
n &=& -2\cos 4\lambda_{\rm ZB} \,. \label{ZBweights}
\end{eqnarray}
The model (\ref{vertices}) with weights (\ref{ZBweights}) is precisely the integrable dilute O($n$) model \cite{Nien1,Nien2,Nien3,Nien4},
here written in the conventions of \cite{ZhouBatchelor}.%
\footnote{The subscript ZB refers to Zhou and Batchelor \cite{ZhouBatchelor}.}
In the following we parameterise the anisotropy by $\lambda$ (spectral parameter) and the loop weight by $\gamma$ (crossing parameter), as
in (\ref{loopweight}). These are linked to the notation \cite{ZhouBatchelor} through the relations
\begin{eqnarray}
 \lambda_{\rm ZB} &=& {\pi \over 2} - {\gamma \over 2} \,, \nonumber \\ 
 u_{\rm ZB} &=& \mathrm{i}\lambda \,.
\end{eqnarray}

In all the following we also rescale $\rho_4$ and $\rho_5$ by a factor $-1$. As noticed in \cite{Nien1} this simple gauge transformation changes
neither the partition function nor the correlation functions, and it has the advantage of rendering all the Boltzmann weights at the polymer point
positive, as shown explicitly in (\ref{polymerweights}). This is an important point to us, since we want to show that this dilute polymer model
has a non-compact continuum limit (i.e., with a continuous spectrum of critical exponents), although the only non-unitary ingredient stems from the
non-locality of the loop weight $n$. This should be compared in particular with the earlier examples of statistical mechanics models with a
non-compact continuum limit, that were defined in terms of negative Boltzmann weights \cite{IkhlefAFPotts,IkhlefBH,VernierPotts}.

With these conventions, the weights (\ref{ZBweights}) admit two isotropic points which are
\begin{equation}
 \lambda_\pm = \mathrm{i}\left(\frac{3 \gamma}{4} \pm \frac{\pi}{4}\right) \,.
\end{equation}
and for each of those there are two polymer ($n=0$) points in the fundamental interval $\gamma \in [0,\pi]$, namely $\gamma = \frac{\pi}{4}$
and $\gamma = \frac{3 \pi}{4}$.
The correspondence with the three regimes of Ref.~\cite{Nien4} and the four branches of Ref.~\cite{Nien1} is as follows.
With $u = u_+$, we are in regime I, corresponding to dense polymers (branch 1) for $\gamma = \frac{\pi}{4}$,
and to dilute polymers (branch 2) for $\gamma = \frac{3 \pi}{4}$. Explicitly this leads to the weights
\begin{eqnarray}
 p_{\rm den} &=& \sqrt{2} \cos\frac{\pi}{16} \approx 1.38704 \,, \nonumber \\
 K_{\rm den} &=& \frac{1}{\sqrt{2}\cos\frac{\pi}{16}-2\sin\frac{\pi}{16}} \approx 1.00315 \,, \nonumber \\
 \tau_{\rm den} &=& 1+\frac{1}{\sqrt{2}} - \sqrt{\frac{1}{2}+\frac{1}{2\sqrt{2}}} \approx 0.783227 \label{denseweights}
\end{eqnarray}
and
\begin{eqnarray}
 p_{\rm dil} &=& \sqrt{2}\sin\frac{3\pi}{16} \approx 0.785695 \,, \nonumber \\
 K_{\rm dil} &=& \frac{\sqrt{2}\sin\frac{3\pi}{16}}{1+\cos \frac{\pi}{8}} \approx 0.408391 \,, \nonumber \\
 \tau_{\rm dil} &=&1 - \frac{1}{\sqrt{2}} + \sqrt{ \frac{1}{2}-\frac{1}{2\sqrt{2}}} \approx 0.675577 \,. \label{diluteweights}
\end{eqnarray}
With $u = u_-$, we are in regime III (branch 3) for $\gamma = \frac{\pi}{4}$, corresponding to the point $\Theta_{\rm BN}$
already given in (\ref{polymerweights}), and in regime II (branch 4) for $\gamma = \frac{3 \pi}{4}$ with weights
\begin{eqnarray}
 p_{\rm II} &=& \sqrt{2}\cos{3 \pi \over 16} \approx 1.17588 \,, \nonumber \\
 K_{\rm II} &=&  \frac{4 \sin{3 \pi \over 16}\sin{\pi \over 8}}{1-8\sin\frac{\pi}{8}\sin^2{3 \pi \over 16}} \approx  15.4476 \,, \nonumber \\
 \tau_{\rm II} &=& -\left(1-\sqrt{\frac{1}{2}+\frac{1}{2\sqrt{2}}}\right)\frac{\cos{\pi \over 16}}{\cos {3 \pi \over 16}} \approx - 0.0897902 \,. \label{regime2weights}
\end{eqnarray}
Unlike the other integrable points (\ref{polymerweights}), (\ref{denseweights}) and (\ref{diluteweights}), this latter point (\ref{regime2weights})
does not seem to allow for a gauge transformation that will render all Boltzmann weights explicitly positive.

\section{Transfer matrix spectrum and critical exponents}

\subsection{O($n$) model transfer matrix}

Consider the O($n$) loop model on a square lattice of size $L \times H$ sites in the horizontal and vertical directions respectively, and take periodic boundary
conditions in the former. The corresponding (grand canonical) partition function can be written as 
\begin{equation}
 Z_{L,H} = \sum_{\mathcal{C}} {\rho_1}^{N_1}\cdots {\rho_9}^{N_9} n^{N_{\rm loops}} \,,
 \label{ZLH}
\end{equation}
where the sum is over all possible loop configurations, $N_i$ (with $i=1,\ldots,9$) is the number of occurencies of each type of vertex (\ref{vertices}),
and $N_{\rm loops}$ is number of closed loops within each configuration. Note that both closed contractible loops and loops winding around the
horizontal periodic boundary condition contribute to $N_{\rm loops}$. It is possible to give a different weight to these two loop types by twisting the
model, and we shall make use of this possibility in the sequel.

It is convenient to introduce a transfer matrix formalism \cite{Nien1} in which (\ref{ZLH}) is computed by decomposing the lattice into horizontal slices.
Each configuration of the vertices within the lowermost $t$ horizontal slices induces a {\em connectivity} among the $L$ edge mid-points that form the intersection
of the lattice with a horizontal line cutting the system between row $t$ and row $t+1$. Some of these $L$ points can be empty, and the remaining
points are either connected pairwise via an {\em arc} (a contiguous part of loop situated below the intersection line), or they connect all the way through
the system via a {\em leg} (a contiguous loop segment that sends at the bottom of the lattice).

The set of all possible connectivities for a system of size $L=3$, labeled by the number of legs $\ell$, can be drawn as follows:
\begin{center}
\begin{tikzpicture}[scale=0.7]
\node (l0) at (0.5,1) {$\ell=0$};
\fill[black] (0,0) circle (1pt);
\fill[black] (0.5,0) circle (1pt);
\fill[black] (1,0) circle (1pt);
\fill[black] (0,-1) circle (1pt);
\fill[black] (0.5,-1) circle (1pt);
\fill[black] (1,-1) circle (1pt);
\draw[black, line width=0.4mm, rounded corners=4pt] (0,-1) -- (0.25,-1.25) -- (0.5,-1);
\fill[black] (0,-2) circle (1pt);
\fill[black] (0.5,-2) circle (1pt);
\fill[black] (1,-2) circle (1pt);
\draw[black, line width=0.4mm, rounded corners=4pt] (0.5,-2) -- (0.75,-2.25) -- (1,-2);
\fill[black] (0,-3) circle (1pt);
\fill[black] (0.5,-3) circle (1pt);
\fill[black] (1,-3) circle (1pt);
\draw[black, line width=0.4mm, rounded corners=4pt] (0,-3) -- (0.5,-3.25) -- (1,-3);
\begin{scope}[shift={(5,0)}]
\node (l1) at (0.5,1) {$\ell=1$};
 \fill[black] (0,0) circle (1pt);
\fill[black] (0.5,0) circle (1pt);
\fill[black] (1,0) circle (1pt);
\draw[black, line width=0.4mm] (0,-0.2) -- (0,0.2);
\fill[black] (0,-1) circle (1pt);
\fill[black] (0.5,-1) circle (1pt);
\fill[black] (1,-1) circle (1pt);
\draw[black, line width=0.4mm] (0.5,-1.2) -- (0.5,-0.8);
\fill[black] (0,-2) circle (1pt);
\fill[black] (0.5,-2) circle (1pt);
\fill[black] (1,-2) circle (1pt);
\draw[black, line width=0.4mm] (1,-2.2) -- (1,-1.8);
\begin{scope}[shift={(0,-3)}]
 \fill[black] (0,0) circle (1pt);
\fill[black] (0.5,0) circle (1pt);
\fill[black] (1,0) circle (1pt);
\draw[black, line width=0.4mm] (0,-0.2) -- (0,0.2);
\draw[black, line width=0.4mm, rounded corners=4pt] (0.5,0) -- (0.75,-0.25) -- (1,0);
\fill[black] (0,-1) circle (1pt);
\fill[black] (0.5,-1) circle (1pt);
\fill[black] (1,-1) circle (1pt);
\draw[black, line width=0.4mm] (0.5,-1.2) -- (0.5,-0.8);
\draw[black, line width=0.4mm, rounded corners=4pt] (0,-1) -- (-0.05,-1.125) -- (-0.25,-1.25);
\draw[black, line width=0.4mm, rounded corners=4pt] (1,-1) -- (1.05,-1.125) -- (1.25,-1.25);
\fill[black] (0,-2) circle (1pt);
\fill[black] (0.5,-2) circle (1pt);
\fill[black] (1,-2) circle (1pt);
\draw[black, line width=0.4mm, rounded corners=4pt] (0,-2) -- (0.25,-2.25) -- (0.5,-2);
\draw[black, line width=0.4mm] (1,-2.2) -- (1,-1.8);
\end{scope}
\end{scope}
\begin{scope}[shift={(10,0)}]
\node (l2) at (0.5,1) {$\ell=2$};
 \fill[black] (0,0) circle (1pt);
\fill[black] (0.5,0) circle (1pt);
\fill[black] (1,0) circle (1pt);
\draw[black, line width=0.4mm] (0.5,-0.2) -- (0.5,0.2);
\draw[black, line width=0.4mm] (0,-0.2) -- (0,0.2);
\fill[black] (0,-1) circle (1pt);
\fill[black] (0.5,-1) circle (1pt);
\fill[black] (1,-1) circle (1pt);
\draw[black, line width=0.4mm] (0.5,-1.2) -- (0.5,-0.8);
\draw[black, line width=0.4mm] (1,-1.2) -- (1,-0.8);
\fill[black] (0,-2) circle (1pt);
\fill[black] (0.5,-2) circle (1pt);
\fill[black] (1,-2) circle (1pt);
\draw[black, line width=0.4mm] (1,-2.2) -- (1,-1.8);
\draw[black, line width=0.4mm] (0,-2.2) -- (0,-1.8);
\end{scope}
\begin{scope}[shift={(15,0)}] 
\node (l2) at (0.5,1) {$\ell=3$};
 \fill[black] (0,0) circle (1pt);
\fill[black] (0.5,0) circle (1pt);
\fill[black] (1,0) circle (1pt);
\draw[black, line width=0.4mm] (1.0,-0.2) -- (1.0,0.2);
\draw[black, line width=0.4mm] (0.5,-0.2) -- (0.5,0.2);
\draw[black, line width=0.4mm] (0,-0.2) -- (0,0.2);
\end{scope}
\end{tikzpicture}
\end{center} 
We note that since the vertices (\ref{vertices}) do not allow the crossing of loop segments, the configurations can all be drawn in a non-crossing fashion.
In particular a leg is never trapped inside an arc. The horizontal periodic boundary condition however implies that for $\ell > 0$ some of the arcs will need
to cross the boundary.

From a configuration corresponding to a given set of connectivities at height $H$, the possible configurations at height $H+1$ are obtained by acting
with all the compatible vertices (\ref{vertices}) on each of the $L$ vertices within a row. Each connectivity carries a statistical weight, corresponding
to the Boltzmann weights appearing in the partition function (\ref{ZLH}) for a partially constructed lattice. The addition of the $(H+1)$th horizontal row
can thus be represented by the so-called geometrical (row-to-row) transfer matrix $T^{(L)}$, whose non-zero elements provide the transitions between configurations
at height $H$ and $H+1$. More precisely, the matrix element $T^{(L)}_{ji}$ is equal to the product of Boltzmann weights (\ref{vertices}) for the $L$ vertices
within the $(H+1)$th row---including a factor of $n$ for each closed loop---summed over the diagrams from (\ref{vertices}) that produce connectivity $j$ for the
system of size $H+1$, starting from the connectivity $i$ for the system of size $H$. The transfer
matrix can be depicted as follows
\begin{center}
\begin{tikzpicture}[scale=0.7]
\foreach \x in {0,1,4}
{ 
\draw[black, line width=0.2mm] (\x,0) -- (\x,1);
} 
\draw[black, line width=0.2mm] (-0.5,0.5) -- (1.5,0.5);
\draw[black, line width=0.2mm] (3.5,0.5) -- (4.5,0.5);
\draw[black, line width=0.2mm, dashed, dash pattern=on 2pt off 2pt] (1.5,0.5) -- (3.5,0.5);
\node (0) at (0,-0.75) {$1$};
\node (1) at (1,-0.75) {$2$};
\node (4) at (4,-0.75) {$L$};
\end{tikzpicture}
\end{center}
where each crossing represents the $\check{R}$-matrix of section \ref{section:correspondenceloopvertex} in its loop representation, and it is understood that the
left and right ends of the horizontal line have to be joined to recover the periodic boundary conditions. This is exactly equivalent to equation (2.12) in Ref.~\cite{Nien1}.

It is easily seen that the transfer matrix $T^{(L)}$ has a block-triangular structure, as the number of legs $\ell$ can only be lowered (by contracting one leg with another)
or kept fixed from one row to the next. As will become clear soon all the information we need is encoded in the transfer matrix's eigenspectrum, so we can focus on the
diagonal part, and hence study separately the different subsectors of fixed $\ell$. Note that the transfer matrix is not symmetric, which means it is not necessarily
diagonalisable. It could also be that some of the eigenvalues are complex. However, numerical study shows that, for generic $n$, the matrix is indeed diagonalisable
and all the low-lying eigenvalues (which are those that matter in the scaling limit) are real. For degenerate values of $n$ (such that $q$ is a root of unity), the transfer matrix
is usually not fully diagonalisable, and exhibits Jordan cells \cite{Dubail10,Vasseur11}, even in the low lying sectors, indicating that the corresponding
conformal field theory is logarithmic \cite{Gainutdinov13}. This is, however, a technical detail which does not matter much here, and in general we do not
distinguish genuine and generalised eigenvalues.

Since the model is critical, this geometrical transfer matrix has a large number of very close eigenvalues $\Lambda_\alpha$, whose scaling with the size of the system
is related with critical exponents, following the usual conformal invariance prediction. Calling $\Lambda_0^{(L)}$ the largest of these, and $\Lambda_\alpha^{(L)}$ any
of the following eigenvalues,%
\footnote{It is convenient to view our two-dimensional statistical mechanics model as a one-dimensional quantum model (spin chain) evolving in imaginary time.
In this picture the leading and next-to-leading transfer matrix eigenlevels are the ground state and low-lying excitations of the corresponding quantum Hamiltonian.
We shall frequently refer to this equivalent point of view in the following.}
we have \cite{Cardy86,Cardy84}
\begin{eqnarray}
 -\frac{\log \Lambda_0^{(L)}}{L} &=& f_{\infty} - \frac{\pi c }{6 L^2} + o(L^{-2}) \,, \nonumber \\
 -\frac{\log \Lambda_\alpha^{(L)}}{L}+\frac{\log \Lambda_0^{(L)}}{L} &=&  \frac{2 \pi x_\alpha }{ L^2} + o(L^{-2}) \,,
 \label{Lalpha}
\end{eqnarray}
where $f_{\infty}$ is the free energy per vertex of the infinite system, $c$ is called the central charge,
and $x_\alpha  = \Delta_{\alpha}+ \bar{\Delta}_{\alpha}$, where $({\Delta}_{\alpha},\bar{\Delta}_{\alpha})$ are the holomorphic and antiholomorphic conformal weights
of the operator associated with the state $\alpha$. The central charge and set of conformal weights associated with the leading transfer matrix eigenlevels
encode the whole set of critical exponents, as will be illustrated in the following sections. 

\subsection{Relation with the IK transfer matrix, and exact results}

In order to investigate the critical content at the $\Theta_{\rm BN}$ point, our next task is therefore to identify the leading eigenlevels of the O($n$) model
transfer matrix, or alternatively, those of the Izergin-Korepin model transfer matrix. The latter is defined similarly in terms of the $\check{R}^{(2)}$ matrix
(\ref{R2}), and it acts in the product $\left(\mathbb{C}^3\right)^{\otimes L}$ of spin-1 representations (see \cite{VJS1} for more details). It commutes with
the total magnetisation 
\begin{equation}
 m= \sum_{i=1}^{L}S^{z}_{i} \,,
\end{equation}
which is the vertex model counterpart of the geometrical quantum number $\ell$ mentioned earlier (see also below). 
A crucial point is to control the relationship between the loop model and the different gauge transformed variants of the vertex (Izergin-Korepin) model.
There are well-defined procedures to do so, based either on algebraic considerations \cite{PS}, or on explicit calculations of generating functions of
levels \cite{RS01}. Note that this is not only a technical question: as the loop model is essentially non-local, the kind of question---typically, a geometrical
correlation---one wants to answer will affect the details, or even the nature, of the correspondence. In what follows, we restrict ourselves to the eigenvalues
of the `natural' polymer transfer matrix, discussed in \cite{Nien1}: it encodes in particular all the exponents we are interested in, and that have been studied
in the literature.  As a matter of fact, the precise relationship between the O($n$) and Izergin-Korepin eigenvalues depends crucially depends on the number
of through-lines $\ell$: 
\begin{itemize}
\item In sectors $\ell > 0$, the eigenvalues of the loop transfer matrix are found in the spectrum of the vertex transfer matrix with periodic boundary conditions,    
\begin{equation}
 T^{(L)} = \mathrm{Tr}_h \left(\check{R}^{(2)}_{h1}\check{R}^{(2)}_{h2}\ldots \check{R}^{(2)}_{hL} \right) \,,
 \end{equation}
in the sector of magnetisation $m= \ell$ (or, equivalently, $m=-\ell$). Here the trace is taken over the horizontal space $h = \mathbb{C}^3$. 

\item Things are slightly different in the sector $\ell=0$ where the loop model allows for noncontractible loops, that is, closed loops that wrap around the
cylinder. These must be weighted $n$---the same as the contractible loops---which requires taking for the vertex model the following {\it twisted} transfer matrix    
\begin{equation}
  T^{(L)}_{\rm vertex} = \mathrm{Tr}_h \left(\check{R}^{(2)}_{h1}\check{R}^{(2)}_{h2}\ldots \check{R}^{(2)}_{hL} \mathrm{e}^{\mathrm{i} \varphi S_z} \right)
 \end{equation}
in the sector of zero magnetisation, $m= 0$. This transfer matrix differs from the periodic one by the presence of a boundary twist term
$\mathrm{e}^{\mathrm{i} \varphi S_z}$, where $S_z$ is now the magnetisation along the horizontal (or auxiliary) space $h$. Summing over the two possible
orientations of a non-contractible loop yields a weight $\mathrm{e}^{\mathrm{i}\varphi} + \mathrm{e}^{-\mathrm{i}\varphi}$, which can be made equal to $n$
of Eq.~(\ref{loopweight}) by taking $\varphi = \pi - 2 \gamma$. The eigenvalues of the loop model transfer matrix are a subset of those of the vertex model
with this particular value of the twist. 
\end{itemize}

Let us again recall that the Izergin-Korepin model exhibits several regimes. We find it convenient, in order to identify these regimes, to restrict to
$\gamma\in [0,\pi]$, and use the symmetries of the weights. The model (\ref{R2}) has two particular values of the spectral parameter,
$x=\pm \mathrm{i} q^{3 \over 2}$, for which the weights are isotropic. The case  $x=\mathrm{i} q^{3 \over 2}$ corresponds to the so-called regime I,
whereas the case $x=-\mathrm{i} q^{3 \over 2}$  splits into regime II  for ${\pi \over 3} < \gamma < \pi$, and regime III for $0 < \gamma < {\pi \over 3}$.
The point $\Theta_{BN}$ that we are interested in corresponds to $x=-\mathrm{i} q^{3 \over 2}$ and $\gamma = {\pi \over 4}$. In the notations of \cite{Nien4},
this is $\Psi={\pi\over 4}$, $\theta=-{\pi\over 4}$. Regime III has been studied in details in \cite{VJS1}, and we will  heavily borrow results from this reference in the following. 

\subsubsection{Structure of the leading transfer matrix eigenlevels}

Regardless of whether we consider the vertex or loop formulation, the transfer matrix commutes with the lattice momentum operator $P^{(L)}$, which is
defined so that $\mathrm{e}^{\mathrm{i}P^{(L)}}$ amounts to a horizontal translation by one lattice unit. The transfer matrix eigenvalues can therefore be
classified according to the value of the momentum, $p^{(L)} = {2 \pi \over L} p$ with $p=0,1,\ldots,L-1$, which in turn is equal to the conformal spin, $\Delta-\bar{\Delta} = p$.%
\footnote{Note that the loop model in the sector $\ell>0$ has more eigenvalues than the vertex model with periodic boundary conditions and magnetisation $m = \pm \ell$.

This is because, for a given $\ell$, one can take each of the $\ell$ through-lines around the axis of the cylinder without affecting the Boltzmann weights, if every
such line acquires in doing so a phase which is an $\ell^{\rm th}$ root of unity. This means that some of the polymer eigenvalues are obtained by taking the vertex
model with $m=\ell$ and twisted boundary conditions, the twist being of the form $e^{2i\pi p\over \ell}$, with $p \wedge \ell=1$ and $p$ integer. The corresponding
critical exponents then satisfy $\Delta-\bar{\Delta}=p+\hbox{integer}$. These exponents are crucial in determining for instance the {\sl winding angle distribution} of the
polymers \cite{DS88}. They are, however, not so important for our purposes, and therefore we will not study them further here.}

From the integrability of the Izergin-Korepin model, the transfer matrix eigenvalues can be computed exactly in terms of a set of complex numbers,
the so-called Bethe roots which are solutions of the Bethe Ansatz equations \cite{VJS1}. Crucially, the number of Bethe roots (and equations) increases
linearly with the size $L$ of the system (more precisely, the number of Bethe roots associated with states in the sector of magnetisation $m$  is $m_1=L-m$),
whereas straight-forward diagonalisation of the transfer matrix would involve solving a set of equations whose number increases exponentially with $L$. 

It is convenient to discuss first the structure of the leading transfer matrix eigenlevels in regime III for the untwisted case (i.e., with periodic boundary conditions).
The ground state (largest eigenvalue) is found in the $m=0$ sector, and corresponds to a root configuration made of a `Fermi sea' of ${L \over 4}$
2-strings (pairs of conjugate roots with imaginary parts close to $\pm \left( {\pi \over 4} - {\gamma \over 4} \right)$). The leading excitations with zero
momentum in the $m=0$ sector are obtained by taking an arbitrary number $j$ of 2-strings off the Fermi sea and replacing them by $j$ pairs of
anticonjugate roots on the axis of imaginary part $\frac{\pi}{2}$. The story is essentially the same in other magnetisation sectors, and from now
on we will therefore label the leading eigenvalues of the transfer matrix by two integers, $(m,j)$. 

We can now extend this to generic values of the twist. There are minor changes in the qualitative description of Bethe root configurations, but the global
structure of eigenlevels unchanged. Going back to (\ref{Lalpha}) we therefore have $\Lambda_0 \equiv \Lambda_{(0,0)}$, and we shall define the critical
exponents $x_{m,j} = \Delta_{m,j}+\bar{\Delta}_{m,j}$ (where we recall that $ \Delta_{m,j}=\bar{\Delta}_{m,j}$ in the zero-momentum sector) or, equivalently,
the effective central charges $c_{m,j} \equiv c - 12 x_{m,j} \equiv c_{0,0} - 12 x_{m,j}$. 

\subsubsection{Sectors with $\ell>0$ ($m\neq0$)}

In the case of non-zero magnetisation, we have already seen that the twist is zero. It follows from our Bethe Ansatz analysis that the leading effective
central charges have the form
\begin{equation}
\frac{ - c_{m,j}}{12} =-\frac{1}{6} + m^2\frac{\gamma}{4 \pi} + \left(N_{m,j}\right)^2 \frac{A(\gamma)}{\left[ B_{m,j}(\gamma) + \log L \right]^2}
\label{eq:ceff_m}
\end{equation}
where $j$ labels the excitations. Here, the $N_{m,j}$ are integers satisfying
\begin{eqnarray}
 N_{m,j} &=& 2j + \frac12 \big( 3 - (-1)^m \big) \,, \label{Nmj_integers}
\end{eqnarray}
and the function $A(\gamma)$ is
\begin{equation}
 A(\gamma) = \frac{5}{2} \frac{\gamma \left(\pi - \gamma \right)}{ \left(\pi - 3\gamma \right)^2} \,.
\end{equation}
The functions $B_{m,j}(\gamma)$ are not known accurately at this stage, but are believed to be universal functions with no $L$ dependence at this order.  

In the limit $L\to\infty$, the $L$-dependent term on the right-hand side of (\ref{eq:ceff_m}) goes to zero at fixed $j$, suggesting that the ground state
in this sector is infinitely degenerate. Although this is, in a sense, true, it is better to interpret what happens by observing that, if one scales $j$ with
$\log L$ so as to keep $s \sim {j \over \log L}$ finite (and continuous), Eq.~(\ref{eq:ceff_m}) can now be interpreted as the signature of a
{\sl continuous spectrum of critical  exponents}.

Such a spectrum is not so familiar in statistical mechanics. It occurs frequently in string theory, where the `targets' of associated conformal field theories
are not compact. This non-compactness leads to continuous spectra of critical exponents just like, in ordinary quantum mechanics, free particles on a
non-compact space (e.g., a real  line) have a continuous spectrum of eigenvalues. In most problems of statistical mechanics, by contrast, these
`targets' are compact. This is well-known in the case of the Coulomb gas representation, for instance, where the bosonic field $\Phi$ is compact,
that is, $\Phi\equiv \Phi+2\pi R$, where $R$ is known as the `compactification radius' \cite{JesperReview}. In the IK model the continuous spectrum can be interpreted
as arising instead from a `non-compact boson', that is, a bosonic degree of freedom $\Phi$ for which, formally, $R=\infty$. More precisely, the $\log L$
dependence in Eq.~(\ref{eq:ceff_m}) can be thought of as arising from an effective compactification radius $R(L)$ that diverges as $L \to \infty$.

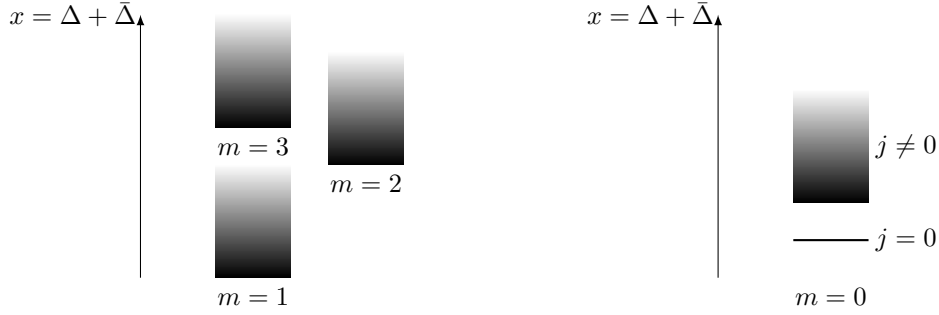
\begin{figure}
\begin{center}
\begin{tikzpicture}[scale=1]
\shade[top color=white, bottom color=black] (0,0) rectangle (1,1.5);
\node[align=center] (1) at (0.5,-0.25) {$m=1$}; 
\shade[top color=white, bottom color=black] (1.5,1.5) rectangle +(1,1.5);
\node[align=center] (2) at (2,1.25) {$m=2$};
\shade[top color=white, bottom color=black] (0,2) rectangle +(1,1.5);
\node[align=center] (3) at (0.5,1.75) {$m=3$}; 
\draw[->,>=latex] (-1,0) -- (-1,3.5);
\node[align=center] (delta) at (-1.9,3.5) {$x = \Delta+ \bar{\Delta}$};
\end{tikzpicture}
\qquad\qquad\qquad
\begin{tikzpicture}[scale=1]
\shade[top color=white, bottom color=black] (0,1) rectangle +(1,1.5);
\draw[thick] (0,0.5) -- (1,.5);
\node[align=center] (0) at (0.5,-0.25) {$m=0$}; 
\node[align=center] (i0) at (1.5,0.5) {$j=0$};
\node[align=center] (i) at (1.5,1.75) {$j\neq 0$};
\draw[->,>=latex] (-1,0) -- (-1,3.5);
\node[align=center] (delta) at (-1.9,3.5) {$x = \Delta+ \bar{\Delta}$};
\end{tikzpicture}
\end{center}
\caption{Structure of the spectrum of critical exponents in the sectors $m>0$ (left), and $m=0$ (right). In the latter case there is a gap between the ground state ($j=0$) and the continuum of $j>0$ excitations, whose magnitude is obtained from (\ref{eq:c00}) and (\ref{eq:c0j}) as $x_g = - \frac{c_{0,1}-c_{0,0}}{12} = \frac{\gamma}{\pi-\gamma}- \frac{1}{4}$.}
\label{fig:spectrum}
\end{figure}

We represent qualitatively the spectrum of critical exponents in the sectors with $m>0$ as shown on the left panel of figure \ref{fig:spectrum}.
Each spectrum has a `socle', above which a continuum starts immediately. We will discuss below what this means for the correlation functions.

\subsubsection{The sector with no through lines ($m=0$)}

The main difference with the case of $\ell,m\neq 0$ is that the leading eigenvalue (which determines the central charge) is separated from a
continuum of critical exponents by a gap. This has to do with the fact that, in the IK model, the whole spectrum contains both continuous
and {\sl discrete states}---just like, for quantum mechanics on the line in the presence of a potential, one can have a spectrum made of both a
continuous and a discrete part. 
The ground state (with $j=0$ and $m=0$) at this value of the twist is given by a discrete state, and leads to the central charge
\begin{equation}
 c_{0,0} = -1 + {12 \gamma^2 \over \pi (\pi - \gamma)} \,.
 \label{eq:c00}
\end{equation}
This coincides with the prediction of Nienhuis {\em et al.}: see Eq.~(7.3c) in \cite{Nien4}, where the notation $\theta = -\gamma$ is used. 
The excited states $j>0$ in the $m=0$ sector form part of the continuum, leading to the effective central charges
\begin{equation}
 c_{0,j} = 2 -{12 \gamma \over \pi} - 12 \left(N_{0,j}\right)^2 \frac{A(\gamma)}{\left[B_{0,j}(\gamma) + \log L \right]^2}
 \label{eq:c0j}
\end{equation}
These observations can be summarised as shown in the right panel of figure \ref{fig:spectrum}.


From all this we can deduce the so-called {\it watermelon exponents} of the loop model. The $m$-leg exponent is defined as 
\begin{equation}
 x_m = -\frac{c_{m,0}-c_{0,0}}{12} \,,
\end{equation}
where the effective central charge is measured at the appropriate twist parameter as detailed above, and the $L\to \infty$ limit
is taken such that the non-compact part does not contribute. We find then
\begin{equation}
 x_m = -{1 \over 4} + {m^2 \gamma \over 4\pi} - \frac{\gamma^2}{\pi(\pi-\gamma)} \,,
\end{equation}
which agrees, once again, with \cite{Nien4}: see Eq.~(7.5) in that reference.

\subsubsection{Leading exponents at the polymer point}

From these results we recover the central charge $c=0$ and the leading watermelon exponents at the polymer point, which
corresponds to $\gamma = {\pi \over 4}$, hence loop weight $n=0$:
\begin{equation}
x_m = {m^2 \over 16} - {1 \over 6} \,.
\end{equation}
This means in particular that 
\begin{equation}
x_1 = -{5\over 48} \,, \qquad x_2 = {1\over 12} \,, \qquad x_4 = {5\over 6} \,.
\end{equation}
This leads, by the usual scaling relations \cite{DS-NPB-87} (namely $\eta = 2 x_1$, $\nu = \frac{1}{2-x_2}$, and $\frac{\gamma}{\nu}=2-\eta$)
to the exponents given in (\ref{exponien}). In addition we identify, like at the usual theta point, the four-leg operator with the thermal operator
coupled to the monomer-monomer attraction energy, which drives the problem away from the (tri)critical point towards the dilute or massive phase.
By the scaling relation $x_4=2-{1\over\nu'}$ one finds $\nu'={6\over 5}$, and thus the crossover exponent
\begin{equation}
 \phi = \frac{\nu}{\nu'} = {10\over 23} \,.
\end{equation}

\section{Physical consequences of non-compactness at the $\Theta_{\rm BN}$ point }

\subsection{Correlation functions}

In the polymer problem, one is typically interested in the correlation functions of observables which are most easily defined in lattice terms,
such as the polymer limit ($n \to 0$) of the spin-spin correlation function. The latter can be expressed as a sum over all configurations 
of (in this case, vertex-interacting) self-avoiding walks joining two given points:
\begin{equation}
G_1(I,J)=\sum_{{\rm VISAW}:\ I\to J} K^{\tiny{\hbox{number of monomers}}}~~ \tau^{\tiny{\hbox{number of doubly visited sites}}}~~ p^{\tiny{\hbox{number of straight segments}}} \,.\label{twoptfct}
\end{equation}
Such correlation function become, at the critical point and in the continuum limit---that is, here, at distances much larger than the lattice
spacing---a complicated sum over correlation functions of conformal fields. The point is that a given lattice observable never corresponds
exactly to a pure scaling field: rather, it can be expanded as an infinite sum over scaling fields, typically of the form
\begin{equation}
{\cal O}_{\tiny{\hbox{latt}}}=C_1\epsilon^{x_1}{\cal O}_1+C_2\epsilon^{x_2}{\cal O}_2+\ldots\label{sumi} \,,
\end{equation}
where the ${\cal O}_i$ are pure scaling fields whose two-point function is normalised to $1/(z\bar{z})^{x_i}$ (for simplicity, we only mention
scalar contributions in this analysis). The analysis of the correlation function
(\ref{twoptfct}) must take into account both this expansion of operators and a similar expansion of the lattice Hamiltonian, which
differs from its continuum limit counterpart also by a sum of irrelevant operators.
In models with a discrete spectrum of critical exponents, this simply leads to
\begin{equation}
G_1(I,J)\propto \left({\epsilon\over r_{IJ}}\right)^{2x_1}+C \left({\epsilon\over r_{IJ}}\right)^{2x_1'}+\ldots\label{sumii} \,,
\end{equation}
where $x'_1>x_1$ is the exponent of the first {\sl corrections to scaling} term.
The discrete spectrum here guarantees that the contributions of the various correction terms to $G_1$ can be well separated at large distance,
and that the asymptotic behaviour is fully determined, at leading order, by $x_1$. This problem is well-known, and has been studied extensively,
for instance for the six-vertex model in \cite{SergeiTerras}.

Meanwhile, when the theory admits a continuous spectrum of critical exponents, one expects sums such as (\ref{sumi})--(\ref{sumii}) to be replaced by integrals:
\begin{equation}
G_1(I,J)\approx \int_{0}^\infty {\rm d}s \, {f_1(s) \over r_{IJ}^{2x_1+2a_1s^2}}\label{corrsum} \,.
\end{equation}
We have here written the continuous part of the spectrum simply as $x_1+a_1s^2$, with $s\sim{i \over \log L}$ and $a_1=4A(\pi/4)$,
by expanding (\ref{eq:ceff_m})--(\ref{Nmj_integers}) to leading order. We could absorb $a_1$ into the definition of the quantum number
$s$, but we refrain from doing so, because $s$ has a precise meaning when the conformal field theory is interpreted in terms of a
non-compact coset sigma model. The amplitude $f_1(s)$ is meanwhile determined by several factors, including the matrix elements between
the lattice spin observable and the eigenstates with quantum numbers $(m,j) = (1,i)$ and the (properly defined) density of states. Even though
the theory is non-unitary, we expect the function $f_1$ to be positive.%
\footnote{This positivity was checked numerically on cylinders of sizes $L=4,6,8,10$, where we explicitly computed the matrix elements
$\langle 1,j | \sum_{i=1}^L S_i^{(+)}| 0,0\rangle $).} 

Interestingly, expressions such as (\ref{corrsum}) have appeared before in statistical mechanics. Most noticeably, at the plateau transition
in the quantum Hall effect for instance, it is expected that the $q^{\rm th}$ moment of the transmission coefficient  between two point
contacts at distance $r$ reads \cite{JMZ}
 $$
 \langle T^q\rangle=\int |\langle Vq,V^*0|\lambda q\rangle|^2 r^{-2\Delta_\lambda}\mu(\lambda) \, {\rm d}\lambda \,,
 $$
where the scalar products and the measure $\mu(\lambda)$ are known, while the exponents $\Delta_\lambda$ only have a conjectured form. In this case as
well, the presence of the integral is ultimately related with the non-compactness of the target. 
 
The practical point of an expression such as (\ref{corrsum}) is that the term going as $r^{-2x_1}$---which is the only one identified by the
naive determination of the one-leg operator---predicts the leading behaviour of the correlation function only at extremely large $r$. It is difficult
to say much more without having a better idea of the function $f_1$, but it is nonetheless clear from the outset, that any attempt to match
an expression such as (\ref{corrsum}) with a pure power law $r^{-2x}$ will lead to an effective exponent $x>x_1$. In the specific context of
numerical simulations, it is of course well-known that the results based on the study of a system of linear size $L$ provide only an effective
exponent $x_1(L)$, which will have to be extrapolated by using finite-size scaling. Our point is that such extrapolations are usually thought
to contain only power law corrections, and this does not correctly take into account the presence of $\log L$ in (\ref{eq:ceff_m}).

Writing now
\begin{equation}
 G_1(r)=r^{-2x_1}\int_0^\infty {\rm d}s \, f_1(s)e^{-2a_1s^2 \ln r} \,,
\end{equation}
there are known techniques \cite{Erdelyi} to  extract an asymptotic expansion for $G_1$ based on the analyticity properties of $f_1$ near the origin. We will restrict ourselves to the leading term here, and simply observe that if 
 \begin{equation}
 f_1(s)=s^{z}\left[b_0+sb_1+\ldots\right] \,, \quad \mbox{for } s\to 0 \,, \label{f1s}
 \end{equation}
 we will have 
 \begin{equation}
 G_1(r)= r^{-2x_1}(\ln r)^{-(1+z)/2}\left[c_0+{c_1\over\sqrt{\ln r}}+\ldots\right] \,.
 \end{equation}

\subsection{Grand canonical Monte-Carlo simulations}
 
Let us compare these predictions with Monte-Carlo sampling of the function $G_1(r)$, both at the point $\Theta_{\rm BN}$ and at the point $\Theta_{\rm DS}$.
At the latter point the peculiar effects of non-compactness are not expected, and indeed it is well-known \cite{DS87} that $x_1 = \frac{1}{4}$.
The polymer configurations are sampled on square lattices of sizes $L \times L$ with periodic boundary conditions in both directions, with
one polymer end fixed at the center. The second end evolves according to an improved version of the backbite algorithm \cite{Backbite1,Backbite2} that
properly takes into account the Boltzmann weight of doubly visited sites and straight segments.%
\footnote{To be precise, the pure backbite algorithm \cite{Backbite1} applies to the fully-packed (Hamiltonian walk) case. It can be adapted to dilute walks
by adjoining grow and retract moves, in a Metropolis scheme, following the last few lines of \cite{Backbite2}. Finally, the possibility of having doubly visited
sites introduces a few additional complications. The most subtle of these concerns the situation where a polymer end point resides precisely at a doubly
visited site, in which case a special backbite-type move is introduced, under which the end point moves infinitesimally across the other loop strand visiting the
concerned site \cite{Backbite3}.}

The trivial, empty configuration which is just represented as a point-like polymer at the center of the $L \times L$ lattice is assigned the winding
numbers $w_x = 0$, $w_y = 0$. Whenever the polymer crosses one vertical or horizontal boundary the corresponding winding number is increased by $\pm 1$. 
We proceed as follows:
\begin{enumerate}
\item Start from the trivial, point-like configuration;
\item Run $500 \times L \times L$ time steps to decorrelate from this initial configuration;
\item Then run $10^7 \times L \times L$ time steps, and sample the configurations with $w_x = w_y = 0$, measuring the end-to-end distance $r$ between the two ends of the polymer. 
\end{enumerate}

This produces a histogram of $r^2$ of the form 
\begin{equation}
H(r) = \sum_{|J-0|^2 = r^2} G_1 (0,J)   + \ldots\,,
\end{equation}
where the dots indicate the finite-size corrections related to the fact that the polymer interacts with itself if part of it goes around the torus in either direction. 
In practice, we produce for each size several such histograms (between $10$ or $20$ depending on the size), and compute the corresponding average and
error bars. The function $G_1(0,J)$ can be extracted from $H(r)$, either by a binning procedure, or for each integer value of $r^2$ by dividing $H(r)$ by
the number of different possible $J$ such that $|J-0|^2 = r^2$, eliminating those for which no such $J$ exist (for instance $r^2 = 3$ cannot be obtained on
the square lattice). 

After normalizing $G_1(0,1) = 1$, for a separation of one lattice spacing, the function we eventually consider is $G_1(r) = G_1(0,r)$, measured for $r$ equal
to a fixed ratio of $L$, that is, $r = \alpha L$ with $\alpha \ll 1$. In practice $\alpha = 0.25$ will turn out convenient. This ensured that with respect to the total
size of the lattice the polymer is almost closing on itself, and therefore the corrections coming from walks going around the torus and back should scale
as $L^{- 2 x_{2}}$, where $x_2$ is the two-leg watermelon exponent, which is positive for both $\Theta_{\rm BN}$ and $\Theta_{\rm DS}$. From the previous
section we expect for large $L$:
\begin{itemize}
\item For $\Theta_{DS}$ (and other usual cases)
 \begin{equation}
\log G_1(\alpha L) = -2 x_1 \log L + A + C \left(\frac{\epsilon}{\alpha L}\right)^{2 (x'_1-x_1)} +  \ldots 
\label{eq:fitG1tDS}
 \end{equation}
 
\item For $\Theta_{BN}$, 
 \begin{equation}
\log G_1(\alpha L) = -2 x_1 \log L + A - \frac{1+z}{2}\log \log L + \frac{c_1}{\sqrt{\log L}} +  \ldots 
\label{eq:fitG1tBN}
 \end{equation}
\end{itemize}

\begin{figure}
\begin{center}
  \includegraphics[scale=0.45]{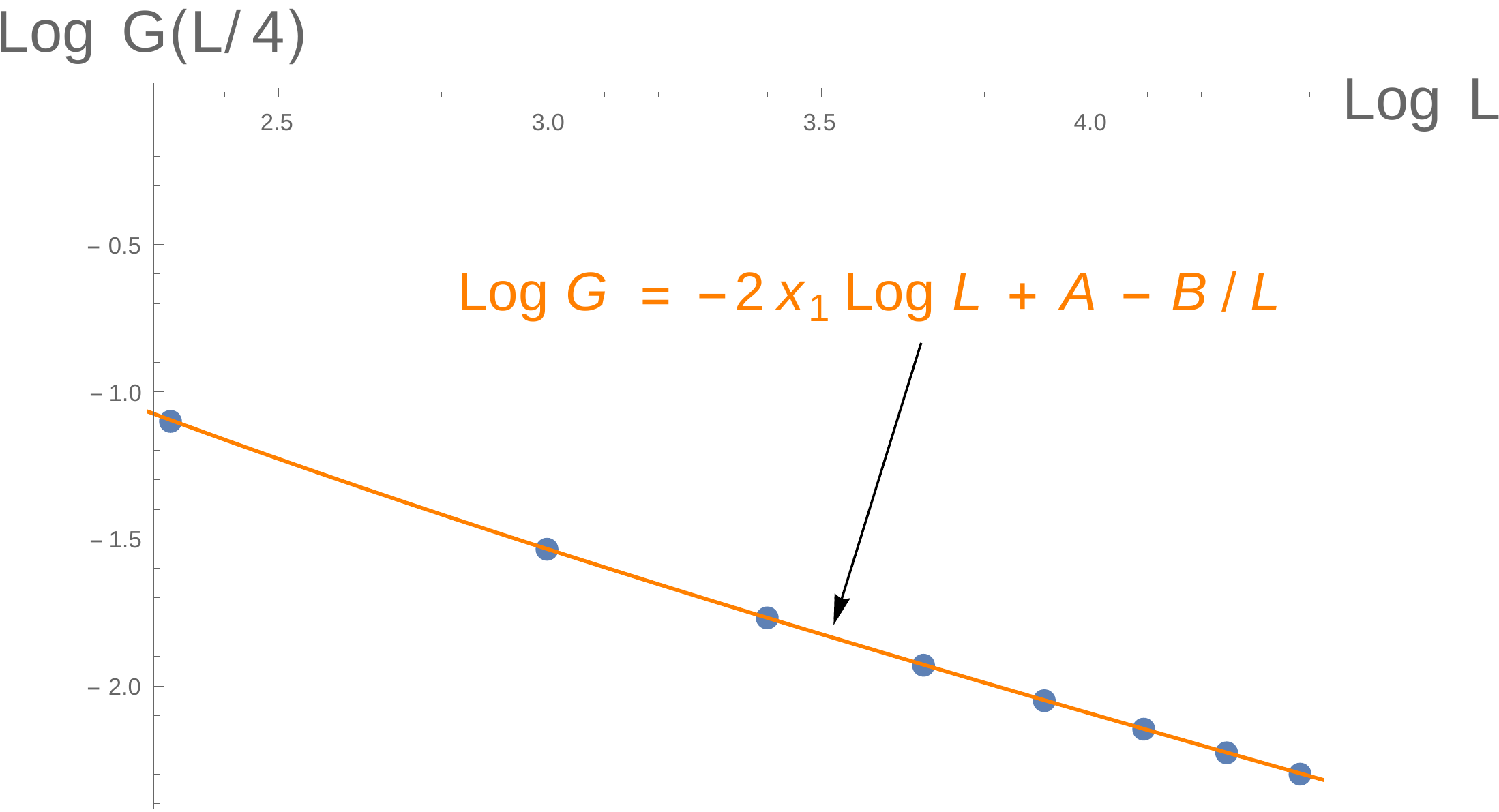}
  
  \vspace{1cm}
 
 \includegraphics[scale=0.4]{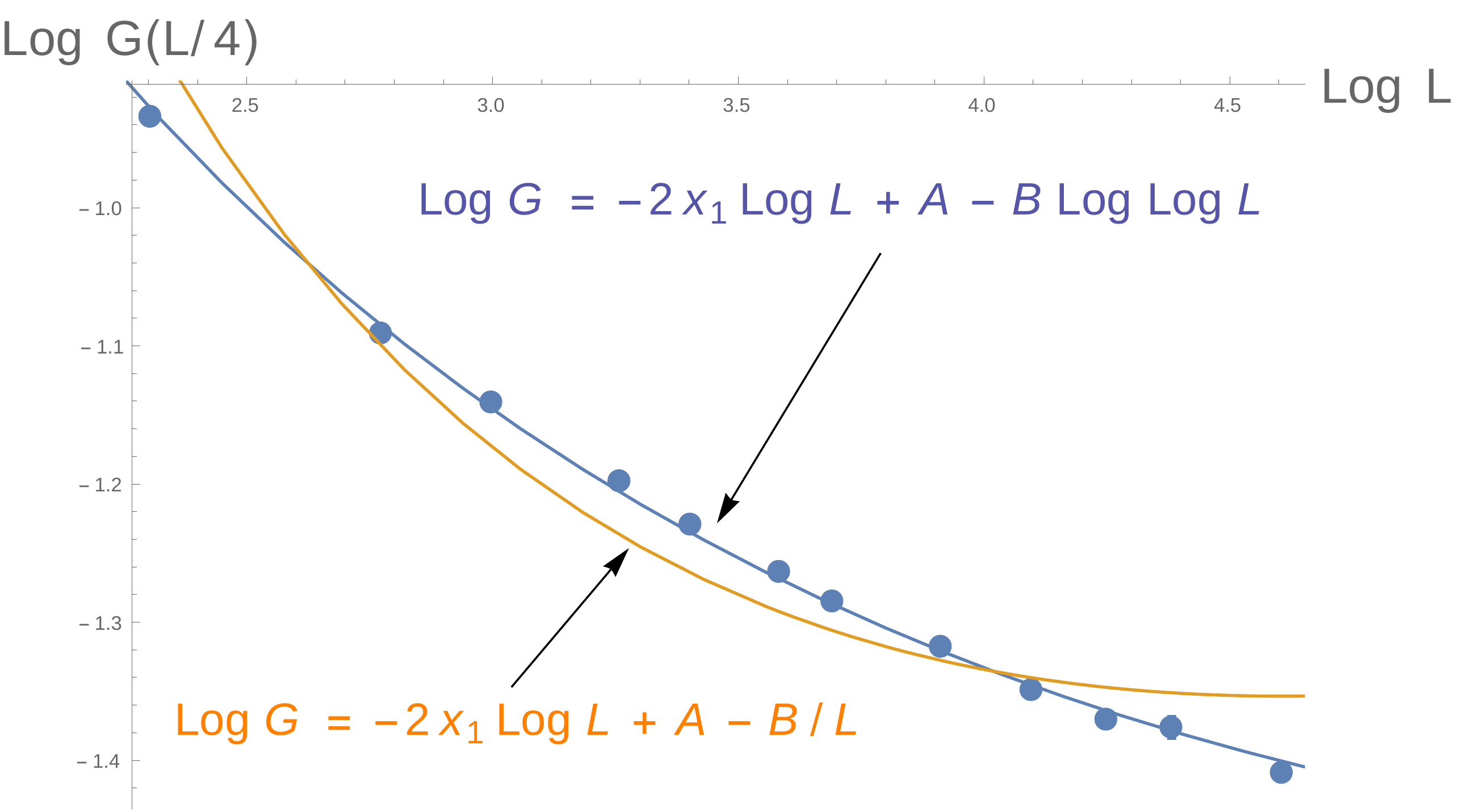}
 \end{center}
\caption{Fit of $\log G_1\left(0.25 L\right)$ against $\log L$, measured from Monte-Carlo sampling at the points $\Theta_{\rm DS}$ (top)
 and $\Theta_{\rm BN}$ (bottom) for $L=10, \ldots,100$ (the error bars are also plotted, but are smaller than the size of the symbols).
 In the first case, a fit of the form (\ref{eq:fitG1tDS}) seems in good agreement with the numerical data, and yields $x_1 \simeq 0.245$.
 In the second case, such a fit is clearly not appropriate (orange curve), and we find much better agreement with (\ref{eq:fitG1tBN}) (blue curve). 
 }
\label{fig:MC_tDStBN}                          
\end{figure} 
 
As shown in figure \ref{fig:MC_tDStBN} (where in the former case we set $C \left(\frac{\epsilon}{\alpha L}\right)^{2 (x'_1-x_1)} \to -\frac{B}{L}$, which takes correct account of the corrections to leading order, while in the latter case $B \equiv \frac{1+z}{2}$), these expectations are nicely corroborated by our results. At $\Theta_{\rm DS}$ we find $x_1 \simeq 0.245$,
which is close to the expected $x_1=\frac{1}{4}$. At $\Theta_{\rm BN}$ we used a fit keeping only the first three terms on the right-hand side
of (\ref{eq:fitG1tBN}), and we plugged in the known value of $x_1 = -\frac{5}{48}$. This yields $z \simeq 1.7$; but note that this estimation is
very sensitive to the presence of higher corrections in the fit, and therefore we shall not trust it too much quantitatively. It is quite striking that, even though
$x_1 < 0$---and therefore at very large distances the function $G_1(r)$ should be an increasing function of $r$---, at the distances that could
be achieved in these simulations $G_1(r)$ is a decreasing function of $r$, leading to an effective exponent $x_1>0$ at intermediate distances. 
 
Albeit the numerical results presented in this section seem rather satisfactory, it could of course well be that $f_1$ has a behaviour quite different
from that predicted by (\ref{f1s}). Unfortunately, at this stage very little is known about this problem.
 
\subsection{Further speculations}
 
In most of the lattice Monte Carlo simulations presented in the literature \cite{Bedini}, the quantity usually studied is however the canonical (fixed length) partition function,  
\begin{equation}
Z_1(I,J)=\sum_{{\rm VISAW}:~I\to J} \tau^{\tiny{\hbox{number of doubly visited sites}}}~~ p^{\tiny{\hbox{number of straight segments}}} \,.
\end{equation}
Even if the function $f_1$ were known, going from $G_1(I,J)$ to $Z_1(I,J)$ would require making hypotheses on the scaling form of the correlation function
in the vicinity of the critical point, $K\neq K_{\rm BN}$, and inverting the corresponding expressions of $G_1$ considered as a Laplace transform of $Z_1$.
There seems little point in speculating about what may happen until more detail is available for the function $f_1$. A short-cut is however provided by the
following argument. Recall the scaling relation
\begin{equation}
{\gamma\over\nu}=2-\eta \,,
\end{equation}
where $\eta=2x_1$. Since the two-point function involves a mix with values of $x_1$ greater than the bottom of the tower, this should lead to estimates
($\nu$ being what it is) for $\gamma$ which are {\sl smaller than the true value}. Indeed, the numerical result \cite{Bedini} reads
$\gamma_{\rm BN}^{num}=1.045<1.152$. Meanwhile, from $x_2=2-{1\over\nu}$ we see that $\nu={1\over 2-x_2}$. Since similarly $x_2$ is decorated
by the tower on top of the minimum value, it will appear larger than its asymptotic value as the polymer length goes to infinity, which means
that $2-x_2$ will appear smaller and thus $\nu$ larger. Indeed, the numerical result \cite{Bedini} is $\nu_{\rm BN}^{num}=0.576>0.522$.
So the discrepancy between numerics and exact results is compatible with our scenario.

\section{The polymer phase diagram}

The main question now concerns the relationship between the theta points $\Theta_{\rm BN}$ and $\Theta_{\rm DS}$, and more generally the nature of the phase diagram for the VISAW. 

It is certainly tempting to consider the  $\Theta_{\rm BN}$ point as an unphysical, infinitely multi-critical point. This is because the spectrum in the sector with no through-lines is continuous above the main gap given by $\Delta_g={c_{0,0}-c_{0,j=1}\over 24}={1\over 24}$. This means there is an {\sl infinity} of relevant operators in the thermal sector, with $\Delta_g\leq \Delta\leq 1$, corresponding formally to an infinity of relevant directions, hence an infinite order of criticality.  On the other hand, it is troubling that the the $\Theta_{\rm BN}$ point seems so natural, and only necessitates the fine-tuning of  a couple of parameters in our phase diagram. Moreover, our experience with the stability of RG fixed point with continuous spectra of  relevant exponents is limited:  one should not be too hasty in deciding that the $\Theta_{\rm BN}$ universality class is unphysical. We must now try to understand this universality class and its vicinity in more detail. 

\subsection{The second virial coefficient}

At $\Theta_{\rm BN}$ ($\gamma={\pi\over 4}$), the ground state of the Izergin-Korepin model with twist $\varphi$ is determined by two possible values of the central charge:
\begin{equation} 
 c(\varphi) = \left \lbrace \begin{array}{ll}
 c^* \equiv 2-{12\varphi^2\over\pi^2} \,, & \mbox{for } \varphi\leq{\pi\over 4} \,, \\[1mm]
 -1+{4\over\pi^2}(\pi-\varphi)^2 \,, & \mbox{for } \varphi\geq {\pi\over 4} \,.
 \end{array} \right. \label{centchform}
\end{equation}
We must take $\varphi={\pi\over 2}$ to make the weight $\tilde{n} = 2 \cos \varphi$ of non-contractible loops equal to zero, which leads to $c=0$.
This is indeed the value expected for the theta point.
It is now interesting to consider what happens near $\varphi={\pi\over 2}$. First, the leading eigenvalue of the transfer matrix takes the form
\begin{equation}
-{\log \Lambda_0^{(L)}\over L}=f_\infty -{\pi c(\varphi)\over 6L^2}+o(L^{-2}) \,,
\end{equation}
where $f_\infty $ is {\sl independent of $\varphi$}, and there are no corrections of order ${1\over L}$ because we use periodic boundary conditions.
When $\varphi\neq {\pi\over 2}$, loops that wind around the axis of the cylinder are allowed, with fugacity $\tilde{n}=2\cos\varphi$. Setting
$\varphi={\pi\over 2}-{\epsilon\over 2}$, this becomes $\tilde{n}=\epsilon+O(\epsilon^3)$. Meanwhile, we can expand the central charge at this order,
and thus the logarithm of the leading eigenvalue:
\begin{equation}
\log \Lambda_0^{(L)}={1\over L}\left({\tilde{n}\over 3}+{\tilde{n}^2\over 6\pi}\right) \,,
\end{equation}
where we have used the natural normalisation where the partition function at $n=\tilde{n}=0$ is unity (hence $f_\infty = 0$). Imagine now taking a long
cylinder of length $L'$: the partition function to leading order will be 
\begin{equation}
\ln Z\approx {L'\over L}\left({\tilde{n}\over 3}+{\tilde{n}^2\over 6\pi}\right) \,.
\end{equation}
Meanwhile, this partition function can be expanded in powers of $\tilde{n}$:
\begin{equation}
\ln Z = \tilde{n} z_1+\tilde{n}^2\left(z_2-{1\over 2}z_1^2\right)+\ldots \,,
\end{equation}
where $z_1$ is the partition function with one non-contractible loop anywhere on the cylinder, and $z_2$ the one with two such loops. Comparing, we find the result
\begin{equation}
z_2-{1\over 2}z_1^2={L'\over L}{1\over 6\pi}\label{partcomb}
\end{equation}
The left hand side combination can be considered as a sort of second virial coefficient \cite{DJ76}. Such terms are not particularly inspiring in general. What is remarkable here is that this difference is {\sl positive}. This is related to the positive curvature
of the second line of (\ref{centchform}) for the central charge as a function of $\varphi$. In models of polymers, such a term is usually negative,
like in the first line of (\ref{centchform}), leading to the opposite sign for the combination of partition functions (\ref{partcomb}). It is easy to understand why such a term is negative for instance for ordinary self avoiding walks: the point is that every term in $z_2$ occurs also in ${1\over 2}z_1^2$, while there are plenty of terms in ${1\over 2}z_1^2$ that are not in
$z_2$---all the terms where two non-contractible loop from each of the $z_1$ have a non-zero overlap. When the model involves bending and attraction, it is not so easy to establish the sign of this term before hand. For instance, there are  terms in $z_2$ where two non-contractible
loops are close to each other, contributing a mutual attraction energy which is not present in the corresponding (geometrically identical) term subtracted in  ${1\over 2}z_1^2$. Nonetheless, in all models known to us---in particular the theta point $\Theta_{\rm DS}$---the combination remains negative. The fact that it is positive in our case indicates the attraction between monomers at the $\Theta_{\rm BN}$ point is unusually strong, a significant fact confirming the belief that $\Theta_{\rm BN}$ is a higher order critical point.

\subsection{Relationship with the dense phase}

To proceed, we make  the simple but crucial observation that the radius of the compact boson for a given value of $\gamma$---that is, for a given value of the
fugacity of contractible loops---in regime III is the {\sl same} as the radius for the compact boson describing the corresponding dense phase of the
vertex or O($n$) model. The dense phase of the vertex model is obtained by increasing the fugacity $K$ of the monomers in the loop version of the
problem, and is in the same universality class as the ``completely packed'' model, which is nothing but the ordinary six-vertex model.  The corresponding
dense phase of the O($n$) model is obtained in the same way, the only difference being that non-contractible loops get fugacity $n$, instead of two for
the vertex model. This observation can be illustrated by considering the polymer case ($n=0$). The $m$-leg (watermelon) exponents are found to be \cite{VJS1}
\begin{equation}
x_m={m^2\over 16}-{1\over 6} \,,
\end{equation}
while they are
\begin{equation}
x_m^{\rm D}={m^2-4\over 16}
\end{equation}
in the case of dense polymers. These two formulas are quite close, since
\begin{equation}
x_m^{\rm D}=x_m-{1\over 12} \,.
\end{equation}
The identical $m^2$ term arises from the fact that both theories involve the same compact boson. The shift of $-{1\over 12}$ occurs because of the
non-compact boson. More precisely, the scaling of the associated eigenvalue in the dense polymer case 
\begin{equation}
-{\log \Lambda_\alpha^{(L)}\over L}=f_\infty+{2\pi \over L^2}\left(x-{c\over 12}\right)
\end{equation}
involves
$$
x_m^{\rm D}-{c^{\rm D} \over 12}={m^2\over 16}-{1\over 12}=x_m^{\rm 1B}-{c^{\rm 1B}\over 12}
$$
with $c^{\rm D} = -2$ for dense polymers, while we have for our critical point
$$
x_m-{c\over 12}={m^2\over 16}-{1\over 6}=x_m^{\rm 1B}-{c^{\rm 2B}\over 12} \,.
$$
Here, $c^{\rm 1B}=1$ denotes the central charge of one boson, while $c^{\rm 2B}=2$ is the central charge of a system of one compact and one
non-compact boson---or the effective central charge of the Black Hole theory. We see therefore that if we took the leading eigenvalue in the $m$-leg
sector and made the non-compact boson massive, the scaling would simply evolve from our model to the dense polymer case:
\begin{equation}
-{\log\Lambda_\alpha^{(L)}\over L}=f_\infty+{2\pi \over L^2}\left(x-{c\over 12}\right)\longrightarrow f_\infty+{2\pi \over L^2}\left(x^{\rm D}-{c^{\rm 1B}\over 12}\right) \,.
\end{equation}
Meanwhile, we also notice that the central charge (that is, the scaling for the ground state energy) is given in our model by a discrete state,
which involves the non-compact boson. Recall indeed the general formula (\ref{centchform}) in our case for a twist $\varphi$.
%
%
We must take $\varphi={\pi\over 2}$ to make the weight of non-contractible loops equal to zero, which leads to $c=0$, while $c^*=-1$. This result
for $c^*$ comes from the central charge $c=1$ of the non-compact boson, and the central charge of the twisted compact boson which must be $c=-2$.
We see that, if we made the non-compact boson massive while leaving the compact one untouched, the discrete level would disappear,  and the value
of the central charge would become $c=-2$ as required for dense polymers. Therefore, in many respects, {\sl the point $\Theta_{BN}$---and more generally,
the critical point in regime III---looks like a dense polymer supplemented by a non-compact degree of freedom.}
\footnote{ It should
nevertheless be borne in mind that the two bosonic degrees of freedom are eventually coupled, leading to more subtle effects.} This confirms the peculiar nature of the $\Theta_{\rm BN}$ critical point, whose features are profoundly different from those expected at an ordinary tricritical point.

\subsection{Probing degrees of freedom}

\begin{figure}
\begin{center}
  \includegraphics[scale=1]{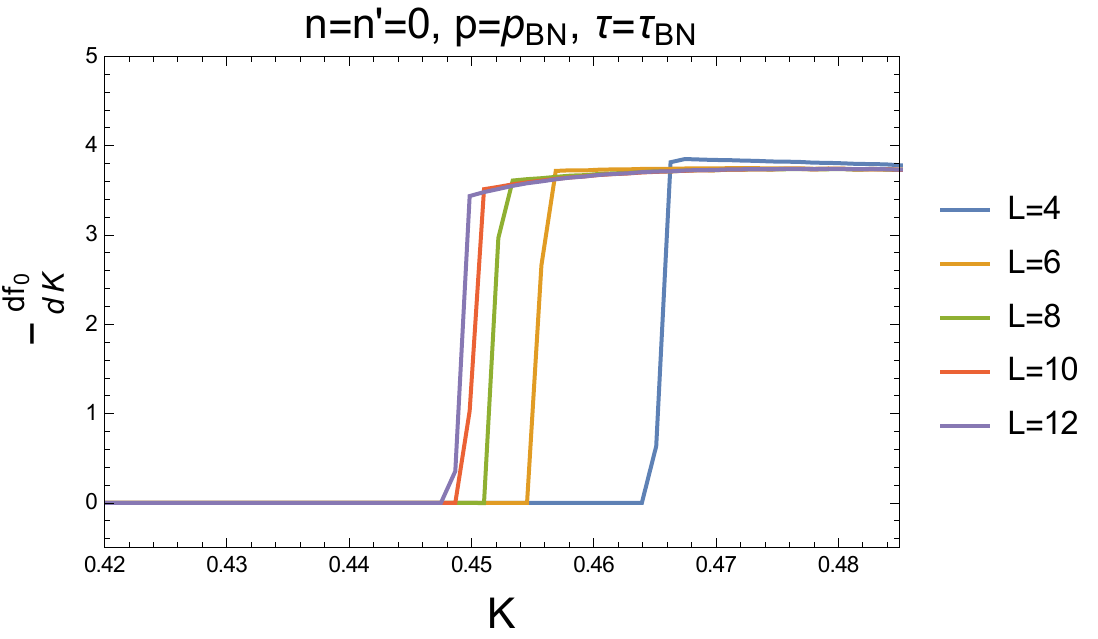}
 \includegraphics[scale=1]{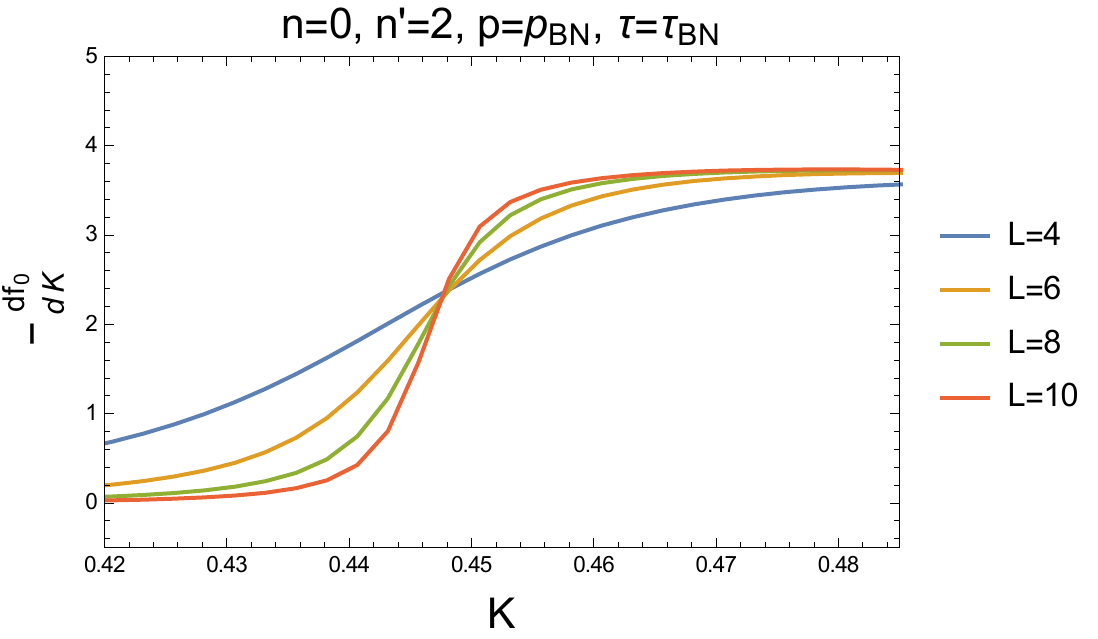}
 \end{center}
\caption{Average density of monomers $- \frac{\partial f_0}{\partial K}$, evaluated numerically from the ground state free energy
 $f_0 = - \frac{1}{L}\log \Lambda_0$, as a function of the monomer fugacity $K$ around the point $\Theta_{\rm BN}$. The top panel
 corresponds to the ``loop case'', where both contractible and non-contractible loops are assigned the same weight $n=0$.
 The bottom panel corresponds to the ``untwisted case'', where non-contractible loops are assigned a weight $\tilde{n}=2$.
 In the former case there is a level crossing at $K \simeq K_{\rm BN}$, which is not present in the latter, but in both cases
  it is apparent that in the limit $L\to \infty$ the density of momomers is discontinous across $K_{\rm BN}$. 
 }                                 
\label{fig:monomerdensity}                          
\end{figure}

It is interesting now  to study numerically the behaviour of the model when the monomer 
fugacity $K$ is increased, $K>K_{\rm BN} \equiv \Theta_{\rm BN}$. In figure \ref{fig:monomerdensity} we have plotted the average density of monomers per lattice edge,
evaluated from the derivative of the (ground state) free energy $f_0$ with respect to the fugacity $K$ (the results are presented both in the untwisted case where non-contractible loops are assigned a weight $\tilde{n}=2$ and in the twisted case where all loops are given the same weight $\tilde{n} = n$).
We find very clear evidence of a {\sl first order phase transition}, with a discontinuity in the monomer density, therefore recovering the conclusions drawn in \cite{FosterDep03} from a corner transfer matrix analysis. In the $K>K_{\rm BN}$ phase,
the fluctuations of density are not critical, and the physics of the model is simply that of the six-vertex model, or ordinary dense loops. 

\begin{figure}
\begin{center}
  \includegraphics[scale=0.8]{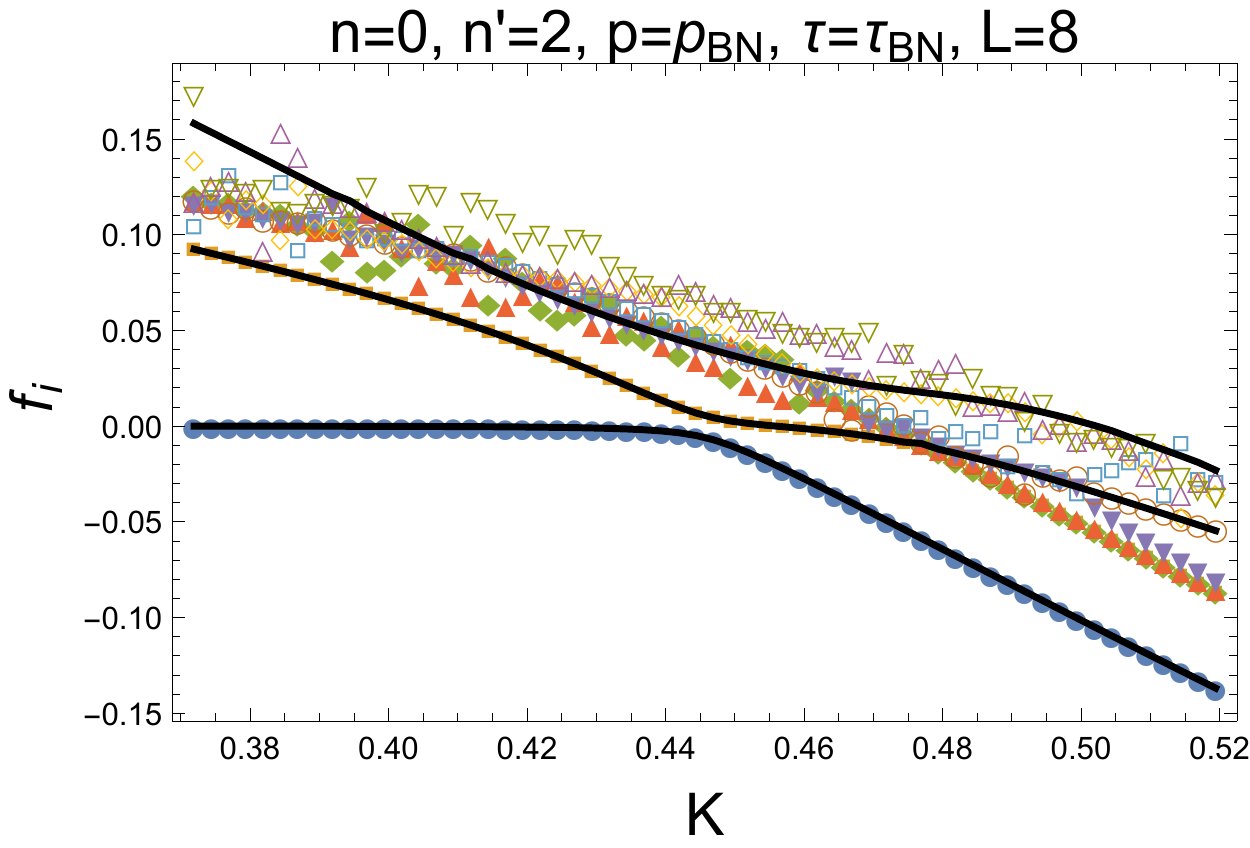}
 \end{center}
\caption{Eigenlevels $f_i = - \frac{1}{L}\log \Lambda_i$ in the sector $\ell=0$, measured for $L=8$ as a function of the monomer fugacity $K$
around the point $\Theta_{BN}$, in the ``untwisted case'', where non-contractible loops are assigned a weight $\tilde{n}=2$. The black continuous
lines are the levels $(0,j)$ (with $j=0,1,2$ from bottom to top). For $j \neq 0$, these levels become massive away from $K=K_{\rm BN}$.
Similar results are observed in other magnetisation sectors and for other values of the weight $\tilde{n}$.
 }                                 
\label{fig:levels}                          
\end{figure}

More generally, we refer to figure \ref{fig:levels}, where it is apparent that the levels with $j \neq 0$ become massive on both the
$K < K_{\rm BN}$ and $K> K_{\rm BN}$ sides of the transition point $\Theta_{\rm BN}$. 

These observations strongly suggest that {\sl the non-compact degree of freedom is associated with fluctuations of monomer density},
since these fluctuations become non-critical precisely when the non-compact degree of freedom disappears. This is, after all, not so surprising.
The emergence of a compact boson in the continuum limit of the six-vertex model follows from the reformulation in terms of a model of heights
defined on the dual lattice, with heights on neighbouring sites defined recursively by the orientation of the arrows separating them. Study of the RG
flow shows that the dynamics of these heights is described by a free bosonic theory. This theory is compactified because, on a torus, the periodicity
of the heights cannot be guaranteed, and the boson much be identified with its shifts by $2\pi R$ \cite{JesperReview}. In {\sl dilute} loop models,
the corresponding vertex model (viz., the IK model) also admits edges that do not carry an arrow. In the regime I of the IK model, these edges
do not have much implication on the continuum limit, as they simply lead to a renormalisation of the  effective coupling constant (radius) of the free boson;
the physics is essentially the same as the one of the dense model. It is however perfectly possible to imagine a scenario where the continuum limit of
the theory is more complicated, and, while the arrow-carrying edges still correspond to height jumps in a height model with a dynamics described by
a free boson at large scales, the arrow-free edges correspond to some additional non-trivial degrees of freedom in the continuum limit. This is in fact
exactly what happens in regime II, where these edges are related with the emergence of an Ising degree of freedom in the continuum limit. 

What we propose now is that the edges that do not carry an arrow---and which correspond, in the loop model, to edges not carrying a monomer---have
a dynamics described in the continuum limit by a non-compact boson. Note that, since the total number of edges on the lattice is of course a constant,
the number of empty edges and number of monomer-carrying edges are related. This means that, in a certain sense, the dynamics of the density at the
critical point of regime III should be related with the non-compact degree of freedom.  


\subsection{A model for $K\approx K_{\rm BN}$}

It is now time to recall a few facts about the Black Hole sigma model which, according to our earlier work \cite{VJS1}, describes the critical $O(n)$ model in regime III. As discussed in the foregoing sections, the question of the {\sl twist} plays an important role. The Black Hole sigma model as discussed, e.g., in \cite{BlackHoleCFT,DVV,RibSch} corresponds, strictly speaking, to the periodic vertex model underlying the $O(n)$ model. In the geometrical  formulation in particular, this means that the non-contractible loops  have weight two. The model where non-contractible loops have weight different from two  is described by a twisted version of the sigma model. For simplicity, we start with the untwisted model, where we recall we have seen the same behaviour---in particular, the possibility of a first order phase transition---as in the twisted one. In the semi-classical limit, the action reads (for simplicity we do not write the dilaton term)
\begin{equation}
A={k\over 4\pi}\int {\rm d}^2x \, {|\partial_\mu \Psi|^2\over 1+|\Psi|^2}
\end{equation}
and describes a sigma model on a target with the shape of a cigar. The compact boson is obtained by considering the model in the limit of
large $|\Psi|$ where the action can be approximated, setting $\Psi=\rho e^{i\theta}$, by 
\begin{equation}
A\approx {k\over 4\pi}\int {\rm d}^2x \, (\partial_\mu\theta)^2 \,.
\end{equation}
The non-compact degree of freedom is, very roughly speaking, the distance along the axis of the cigar, that is, $|\Psi|$. There are some crucial
subtleties here having to do with normalisability, and the best way to understand them is to discuss briefly the mini-superspace approximation \cite{DVV}.
Physically, this is obtained by considering the model on a cylinder of very tiny radius, so that fluctuations in one direction can be neglected,
and the model becomes effectively a zero-dimensional quantum mechanics problem. The Hamiltonian is then nothing but the Laplacian on the target,  
 \begin{equation}
 \Delta=-{2\over k}\left[\partial_r^2+\left(\hbox{coth\,}r+\hbox{tanh\,}r\right)\partial_r+\hbox{coth}^2 r \, \partial_\theta^2\right]  \,,
 \end{equation}
where we have set $\Psi\equiv \sinh r e^{i\theta}$.  There are no $L^2$-normalisable eigenfunctions, but only $\delta$-function normalisable eigenfunctions. They depend on two parameters: one is $n\in \mathbb{Z}$, the angular momentum of rotations around
the axis, and $j=-{1\over 2}+is$ is related with the momentum $s$ along the $r$-direction (the axis) of the cigar. 
 The corresponding eigenvalue of the Laplacian is 
 \begin{equation}
x=\Delta+\bar{\Delta}=-{2j(j+1)\over k}+{n^2\over 2k} \,.\label{minisuperexpo}
 \end{equation}

The full wave functions read (recall that in the mini-superspace approximation, the wave functions coincide with the expressions of the (conformal) fields in terms of the degrees of freedom in the action)
\begin{eqnarray}
 \phi_n^j&=&-{\Gamma^2(-j+{|n|\over 2})\over\Gamma(|n|+1)\Gamma(-2j-1)}e^{in\theta}\nonumber\\
  &&\sinh^{|n|}r F(j+1+{|n|\over 2}, -j+{|n|\over 2}, |n|+1,-\sinh^2 r) \,.
\end{eqnarray}
Setting $j=-{1\over 2}+i{p\over 2}$, and using the notation $\phi_n^p$ for the corresponding eigenfunction, it is possible normalise things in such a way that
\begin{equation}
 \left(\phi_n^p,\phi_n^{p'}\right)=\delta_{nn'}\left[2\pi\delta(p-p')+R_0(p',n)\delta(p+p')\right] \,,
\end{equation}
where the scalar product uses the volume element ${\rm d}v=k\sinh 2r \, {\rm d}r \, {\rm d}\theta$. 

One finds, for the classical reflection amplitude, the expression
\begin{equation}
 R_0(p,n)={\Gamma(ip)\Gamma^2\left({1\over 2}-{ip\over 2}+{n\over 2}\right)\over \Gamma(-ip)\Gamma^2\left({1\over 2}+{ip\over 2}+{n\over 2}\right)} \,.
\end{equation}
A crucial---and general---aspect of these wave functions is that they are {\sl not simple expressions of the $\Psi$ degree of freedom}. Moreover,
while these eigenfunctions have a `normal' behaviour at small $|\Psi|$---namely $\phi_n^j\propto \Psi^n$ for $n>0$ (or $\bar{\Psi}^{-n}$ for $n<0$)---their
behaviour at large $\Psi$ is definitely not usual in quantum field theory. To see this, it is convenient to decompose the eigenfunction as 
\begin{equation}
 \phi_n^p=\phi_{L,n}^p+R_0(p,|n|)\phi_{R,n}^p \,.
\end{equation}
Here,
\begin{eqnarray}
 \phi_{L,n}^p&=&e^{in\theta}(\sinh r)^{-1-ip}F\left({1\over 2}+{ip+n\over 2},{1\over 2}+{ip-n\over 2};1+ip;-{1\over\sinh^2 r}\right) \nonumber\\
& \sim& e^{-r} e^{-ipr+in\theta} \,, \quad \mbox{for } r\to\infty
\end{eqnarray}
and 
\begin{eqnarray}
 \phi_{R,n}^p&=&e^{in\theta}(\sinh r)^{-1+ip}F\left({1\over 2}-{ip+n\over 2},{1\over 2}-{ip-n\over 2};1-ip;-{1\over\sinh^2 r}\right) \nonumber\\
 &\sim& e^{-r} e^{ipr+in\theta} \,, \quad \mbox{for } r\to\infty \,.
\end{eqnarray}
We see therefore that the normalisable eigenfunctions all vanish like $\phi\propto e^{-r}\propto {1\over |\Psi|}$ at large $|\Psi|$. Meanwhile,
eigenfunctions which are polynomial in $\Psi$ all are non-normalisable. Examples include 
\begin{eqnarray}
\phi_0^{j=1}&\propto& (1+2|\Psi|^2) \,, \nonumber\\
\phi_0^{j=2}&\propto& (1+6|\Psi|^2+6|\Psi|^4)
\end{eqnarray}
or
\begin{eqnarray}
\phi_1^{j=1/2}&\propto& \Psi \,, \nonumber\\
\phi_{-1}^{j=1/2}&\propto&\bar{\Psi} \,.
\end{eqnarray}
These `normal' fields all correspond therefore to non-normalisable states, as can easily been checked since %
\begin{equation}
\int {\rm d}^2x  \, |\phi_n^{j \in \mathbb{Z}+1/2}|^2=\infty \,.
\end{equation}

To proceed, we now try to refine the connection between the monomer density and the non-compact degrees of freedom in the CFT.
The operator coupled to $K-K_{\rm BN}$ may be associated with a non-normalisable or with a normalisable state (for a discussion of the state-operator correspondence in CFT with non-compact targets, see, e.g., \cite{Seiberg}). In the first case,
the conformal weight in the $c=2+{6\over k-2}$ theory is negative, so the gap over the $c_{\rm eff}=2$ theory is 
\begin{equation}
\Delta-{c\over 24}=-{j(j+1)\over k-2} -{1\over 4(k-2)}-{2\over 24} \,.
\end{equation}
Such negative values of $\Delta$ would lead to exponents $\nu<{1\over 2}$, and there are no indications of such values in the numerical analysis.
This means that the operator coupled to $K-K_{\rm BN}$ must be normalisable. While there is a continuum of choices (because of the continuous
imaginary part of $j$), it is reasonable, in the absence of other indications, to focus  the bottoms of the continuum, corresponding  to  $j=-{1\over 2}$.
Since monomer pieces involve moreover two links, the most conservative operators one can think of are associated to $j=-1/2, n=0$ and $j=-1/2,n=2$.
From the exact form of the spectrum (beyond the mini superspace approximation (\ref{minisuperexpo})),
\begin{equation}
\Delta(\bar{\Delta})=-{j(j+1)\over k-2}+{(n\pm kw)^2\over 4k} \,,
\end{equation}
we see in particular that operators with $w\neq 0$ have too large a dimension to be relevant, and we do not consider them. 

While the operator with $j=-1/2,n=0$ has conformal weight zero with respect to the effective central charge, it is not necessarily trivial, as can be seen
in the mini-superspace formulation, where it corresponds to a wave function $F(1/2,1/2;1,-|\Psi|^2)$.  We propose therefore that the perturbation
coupled to $K-K_{\rm BN}$ can be represented by the action
\begin{equation}
A\approx {k\over 4\pi}\int {\rm d}^2x \left\{{|\partial_\mu\Psi|^2\over 1+|\Psi|^2}+(K-K_{\rm BN}) \left[\alpha F(1/2,1/2;1,-|\Psi|^2)+\beta (\Psi^2+\bar{\Psi}^2) F(3/2,3/2;3,-|\Psi|^2)\right]\right\}
\end{equation}
with some coefficients $\alpha,\beta$. 
For simplicity let us first look at the physics in the case where $\beta=0$. 

\paragraph{$\beta=0$ case.}

The hypergeometric function is monotonically decreasing from $F=1$ for $\Psi=0$ to $F\sim {1\over |\Psi|}$ when $|\Psi|\to\infty$. If $K<K_{\rm BN}$, we see that
the action is minimised when $F$ is at its maximum, so $|\Psi|=0$. This corresponds to a massive theory, with no critical degrees of freedom left. If $K>K_{\rm BN}$,
the action is minimised when $F$ is as small as possible, so $|\Psi|\to\infty$. For $K<K_{\rm BN}$, the classical action is $A\approx \int {\rm d}^2 x \, (K-K_{\rm BN})\alpha$ 
and for $K>K_{\rm BN}$ it is zero, so we expect a discontinuity of the first-order derivative, that is, a first-order phase transition. 

In the phase $K>K_{\rm BN}$ the fact that $|\Psi|$ is infinite does not mean that the theory is trivial. In fact, this region is precisely the cylinder limit of the cigar,
and one recovers the {\sl same free boson theory} as the compact component of the critical point  $K=K_{\rm BN}$. It is reasonable to consider therefore
the $K>K_{\rm BN}$ phase in our problem as a broken symmetry phase, and the compact boson describing the dense vertex model as a {\sl Goldstone boson}.

In fact, the phenomenology we propose is similar in spirit to the first-order phase transition occurring in systems described by a complex boson, with action 
\begin{equation}
A=\int {\rm d}^2 x \, |\partial_\mu\Psi|^2+(K-K_{\rm BN}) |\Psi|^2(1-|\Psi|^2) \,,
\end{equation}
a model known to describe the superfluid transition (see, e.g., \cite{Minn}).

\paragraph{Generic case.}

More behaviours are possible when $\alpha$ and $\beta$ are non-zero. Of course, the $\alpha$ term is always the most relevant and dominates
whenever $\alpha\neq 0$. When $\alpha=0$ exactly, on the other hand, the phenomenology is quite different from what we see in the $K>K_{\rm BN}$
perturbation. This is because the function $x^2 F(3/2,3/2;3,-x^2)$ has a maximum at a finite value of $x$. Irrespective of the sign of $K-K_{\rm BN}$,
the action is then minimised at this maximum, and for $2\theta=0$ or $\pi$: both degrees of freedom then become massive. Since $\cos2\theta$ is
meanwhile relevant, it is very likely that this scenario persists in the presence of fluctuations, and does not lead to the observed first-order phase transition.

\bigskip

We conclude from this discussion that the transition when $K$ crosses $K_{\rm BN}$ in the untwisted model is driven by a field of dimension $0$ with respect to the effective
central charge $c_{\rm eff}=2$, and of dimension $\Delta={1\over 4(k-2)}$ with respect to the central charge of the Black Hole theory. Unfortunately, the literature
on perturbed field theories having non-normalisable ground states is lacunary, and we are not sure how these values of conformal weights are related
with the scaling of physical quantities as $K$ crosses $K_{\rm BN}$. Certainly, both $\Delta=0$ and $\Delta={1\over 4(k-2)}$ are very small in the domain of
values of $k$ reachable in regime III, leading to values of $\nu$ very close to $\nu={1\over 2}$. The accuracy of our numerical calculations did not allow
us to determine the possible difference $\nu-{1\over 2}$ with reasonable accuracy. 

Turning now to the loop model, it is generally believed---and we have checked this numerically---that the ground-state energy of the theory (or the free
energy per site in the 2D statistical point of view) is independent of the boundary conditions. This means that the scaling should be described by the
operator with weight $\Delta={1\over 4(k-2)}$ again, unless---as happens in the case of the six-vertex model and its loop formulation for instance---the change
of boundary conditions leads to the cancellation of some terms, and another, related but different, scaling dimension (see, e.g., \cite{ResSal} for a discussion of this point).
In our case however, the corresponding value 
\begin{equation}
\nu={2(k-2)\over 4k-9}
\end{equation}
seems to fit numerics reasonably well. Moreover, it is definitely the correct value for $k=8, n=0$. In this case indeed, one can directly identify the operator
coupled to $K-K_{\rm BN}$ with the two-leg operator, well-known in polymer theory, and whose dimension is $\Delta={1\over 24}$ at this point. We conclude with the natural result that, to lowest order in the density, we have indeed that the monomer density is proportional to $|\Psi|^2$. 

\subsection{The multicritical line}

The analysis of the perturbations from the Black Hole CFT at $\Theta_{\rm BN}$ into the dense and massive phases can, at first sight, be extended to the perturbation along the multicritical line
(the violet line in the phase diagram of figure \ref{fig:phasediagram}). But while it is very natural to expect that this perturbation is relevant (and thus that there is a flow away from $\Theta_{\rm BN}$ in all directions of the phase diagram in figure \ref{fig:phasediagram}), the numerical evidence for this scenario is disappointing.

Of course,  investigating the critical behaviour along the multicritical line is a very complicated task. The precise location of the multicritical points away
from  $\Theta_{\rm BN}$ and $\Theta_{\rm DS}$ is unknown, and numerical estimations of critical exponents are plagued by
the flow towards the dense or massive phases, as well as the proximity with the Ising or dilute critical surfaces. At this stage, we feel that an exploration of the phase diagram with conclusive numerical results is beyond the scope of this work. Here, we will content ourselves by investigating a much simpler question, namely 
what happens to the non-compact boson precisely at $\Theta_{\rm DS}$. This can be done  by following continuously the levels $(m,j)$ from $\Theta_{\rm BN}$. Moreover, we restrict for simplicity to the twisted case, where non-contractible loops have weight $n$. 

Before doing so, it is useful to recall a few results about the physics at the point $\Theta_{\rm DS}$. The latter belongs to a more general integrable
line labeled `branch 0' in \cite{Nien1}, and which was solved exactly by Bethe Ansatz in \cite{Batch}. Most of the results can actually be derived from
the observation (reviewed above in section~\ref{sec:2}) that the configurations of the model with weights
$(n, K_{\rm DS}, p_{\rm DS}, \tau_{\rm DS}) = (n,\frac12,0,2)$ can be mapped onto
the configurations of completely packed loops with weights $\widetilde{n} = n + 1$, or equivalently, onto a $Q = (n+1)^2$ state Potts model. The
latter model is critical for $Q \le 4$, and well understood from a Coulomb gas construction \cite{JesperReview}: the continuum limit is described
by a compact boson, and parametrizing $\widetilde{n} = -2 \cos \pi g$ the central charge reads 
\begin{equation}
c = 1 - 6 \frac{e_0^2}{g} \,,
\end{equation}
where $e_0$ is related to the weight $\widetilde{n}_{\rm nc}$ of non-contractible completely packed loops by $\widetilde{n}_{\rm nc} = 2 \cos \pi e_0$. In particular, for the
`pure loop' case where contractible and non-contractible loops have the same weight, $e_0 = 1-g$ so that
\begin{equation}
c=1-6\frac{(1-g)^2}{g} \,.
\end{equation} 
The point $\Theta_{\rm DS}$ corresponds to $n=0$, that is $g = \frac{2}{3}$, and so $c=0$.  
Turning to the critical exponents, they read 
\begin{equation}
x_{e,m} = \frac{e(e-e_0)}{2 g} + \frac{g m^2}{2} \,,
\label{eq:CGexponents}
\end{equation}
that is, at $\Theta_{\rm DS}$, 
\begin{equation}
 x_{e,m} = \frac{e(3e-1)}{4} + \frac{m^2}{3} \,.
 \label{eq:CGTDS}
\end{equation}

We now come back to the levels $(m,j)$. Contrarily to what was observed in the case of $K$ perturbations, these appear to remain critical at $\Theta_{\rm DS}$.
In particular, we have studied in detail the first few levels $(0,j)$ in the sector of zero magnetisation. The following observations hold:
\begin{itemize}
\item $(0,0)$ is still the ground state at $\Theta_{\rm DS}$ (there is no level crossing);
\item $(0,1)$ has at $\Theta_{\rm DS}$ the conformal weight $\Delta^{\rm DS}_{0,1} = \frac{g}{2}$;
\item $(0,2)$ has at $\Theta_{\rm DS}$ the conformal weight $\Delta^{\rm DS}_{0,2} = 2g$.
\end{itemize}
From these observations it is quite natural to conjecture that the levels $(0,j)$ become degenerate with the magnetic excitations of the compact boson
at $\Theta_{\rm DS}$, namely 
\begin{equation}
\Delta^{\rm DS}_{0,j} = x_{0,j} \,.
\end{equation}

It is easy to imagine scenarios where the non-compact boson becomes massive, and the compact boson keeps its radius (as in the flow to the dense phase), or acquires a new value of the radius. It is also possible to imagine scenarios involving the emergence of Ising degrees of freedom, as observed on the blue surface in figure  \ref{fig:phasediagram}---this would simply correspond, in the complex field picture, to the introduction of an additional gauge field \cite{Foda}. But we cannot, for now, imagine a simple scenario connecting highly degenerate conformal weights at the $\Theta_{\rm DS}$ point to the continuum of weights at the $\Theta_{\rm BN}$ point. This will have to wait for further studies. 
Incidentally, this will require, in particular, furthering our understanding of the physics in the $p=0$ plane \cite{p0unpub}.

\section{Conclusion}

In this paper we have partly unraveled the mystery of the theta point in two dimensions. We have established that the BN point found in \cite{Nien1} is very peculiar
from the point of view of critical systems: its correlation functions do not decay as discrete sums of power laws at large distance, but involve instead a
{\sl continuum of critical exponents}. This is expected to lead to extremely strong corrections to scaling, and explains why the results of numerical simulations
have always been `off' compared to the well-established theoretical values. We have also seen that the BN point is {\sl extremely unstable}, and represents,
in a certain sense, an infinite order of criticality. We note that these features are somewhat similar to the ones of the theta point for interacting trails. As discussed in \cite{Nahum}, this point is described also by a non-compact CFT, albeit of a rather trivial type (a Goldstone phase sigma model, free at low energy \cite{JRS03}), with trivial critical exponents. 

One question that we have unfortunately not been able to fully answer is the role of the $\Theta_{\rm BN}$ point in the phase diagram of polymers. We still believe that the point $\Theta_{\rm DS}$ is the  tricritical point for generic  lattices 
and short distance interactions. It may be that it is {\sl also} the generic tricritical point for the model we discuss in this paper, where loops osculate at the vertices (with weight $\tau$, see figure~\ref{fig:VISAWconfig}). This is what conventional thinking would suggest, since the critical theory at the $\Theta_{\rm BN}$ point involves an infinity of relevant operators. Moreover, it is certainly difficult to imagine a flow from $\Theta_{\rm DS}$---essentially described by a single boson---to $\Theta_{\rm BN}$, which involves so many more degrees of freedom. On the other hand, we have not seen convincing numerical evidence for this flow. On top of this, it turns out that, for the model  in figure \ref{fig:VISAWconfig}, the solvable point with universality class of $\Theta_{\rm DS}$ enjoys an additional symmetry, which is broken as soon as `straight segments' are allowed: the number of monomers on the even sub lattice minus the number of monomers on the odd sub lattice is indeed conserved by the interactions. This means, in fact, that the model enjoys an underlying $SU(n)$ symmetry, larger than the $O(n)$ symmetry present in the rest of the phase diagram \cite{p0unpub}. From this point of view, it is tempting to expect that there is an RG flow towards $\Theta_{\rm BN}$, which was suggested in the original paper \cite{Nien1}. But we have not seen clear numerical evidence for this either. It could be that there is a---yet unidentified---intermediate fixed point making these observations compatible. We hope to be able to report on this soon.

\subsection*{Acknowledgements} 

We thank A.D.\ Sokal and  B.\ Nienhuis for discussions. We also thank Y.\ Ikhlef for his collaboration on  \cite{Backbite3}.
Support from the Agence Nationale de la Recherche (grant ANR-10-BLAN-0414: DIME) and the Institut Universitaire de France is gratefully acknowledged.

\end{document}